\documentclass[fleqn,usenatbib]{mnras}
\usepackage[utf8]{inputenc}
\usepackage[T1]{fontenc}
\usepackage{graphicx}
\graphicspath{ {figs/} }
\usepackage{amssymb}
\usepackage{gensymb}
\usepackage{amsmath}
\usepackage{bm}
\usepackage{lineno}
\usepackage{booktabs}
\usepackage{multirow}
\usepackage{color}
\usepackage{xcolor}

\usepackage{mhchem}
\usepackage{tikz}
\usetikzlibrary{shapes, arrows.meta, positioning}

\newcommand{\coreeff}{$\chi^{\rm Fe}_{\rm m}$}

\title[Rocky planet water inventories]{Mantle mineralogy limits to rocky planet water inventories}

\author[C. M. Guimond et al.]{
Claire Marie Guimond,$^{1}$\thanks{E-mail: cmg76@cam.ac.uk (CMG)}
Oliver Shorttle,$^{1,2}$
John F. Rudge$^{1}$
\\
$^{1}$Department of Earth Sciences, University of Cambridge, Downing Street, Cambridge CB2 3EQ, UK\\
$^{2}$Institute of Astronomy, University of Cambridge, Madingley Road, Cambridge CB3 0HA, UK
}

\date{Accepted XXX. Received YYY; in original form ZZZ}

\pubyear{2022}

\begin{document}
\label{firstpage}
\pagerange{\pageref{firstpage}--\pageref{lastpage}}
\maketitle

\begin{abstract}
Nominally anhydrous minerals in rocky planet mantles can sequester multiple Earth-oceans' worth of water. Mantle water storage capacities therefore provide an important constraint on planet water inventories. Here we predict silicate mantle water capacities from the thermodynamically-limited solubility of water in their constituent minerals. We report the variability of upper mantle and bulk mantle water capacities due to (i) host star refractory element abundances that set mantle mineralogy, (ii) realistic mantle temperature scenarios, and (iii) planet mass. We find that transition zone minerals almost unfailingly dominate the water capacity of the mantle for planets of up to $\sim$1.5 Earth masses, possibly creating a bottleneck to deep water transport, although the transition zone water capacity discontinuity is less pronounced at lower Mg/Si. The pressure of the ringwoodite-perovskite phase boundary defining the lower mantle is roughly constant, so the contribution of the upper mantle reservoir becomes less important for larger planets. If perovskite and postperovskite are relatively dry, then increasingly massive rocky planets would have increasingly smaller fractional interior water capacities. In practice, our results represent initial water concentration profiles in planetary mantles where their primordial magma oceans are water-saturated. This work is a step towards understanding planetary deep water cycling, thermal evolution as mediated by rheology and melting, and the frequency of ocean planets.

\end{abstract}


\section{Introduction}

Water has a major effect on planetary processes, from their deepest interiors to their surfaces and atmospheres. In a planet's atmosphere, water is likely to be both the major greenhouse gas species and source of clouds, critical therefore for climate and atmospheric dynamics \citep[e.g.,][]{pierrehumbert_thermostats_1995,frierson_gray-radiation_2006,frierson_gray-radiation_2007}. At the surface, the amount of water trades off with topography to determine whether or not continents are exposed \citep{cowan_water_2014, honing_continental_2016, guimond_blue_2022}. In extreme cases, planets with high surface water inventories will be ocean worlds, presenting possible challenges for volatile cycling \citep{kitzmann_unstable_2015, noack_water-rich_2016, nakayama_runaway_2019, honing_carbon_2019, krissansen-totton_waterworlds_2021}, prebiotic chemistry \citep{patel_common_2015, rimmer_origin_2018}, and the detection of any biotic \ce{O2} \citep{glaser_detectability_2020, krissansen-totton_oxygen_2021}. Even water sequestered far below the surface can have a profound influence on planetary evolution: water changes the mantle rheology (i.e., viscosity) and so the thermal evolution of the interior \citep{karato_rheology_1993, seales_deep_2020}; it facilitates mantle melting by lowering the mantle solidus and so promotes volcanism \citep{green_experimental_1973, katz_new_2003}; and it may be responsible for the initiation of plate tectonics and the birth of continents on Earth \citep{korenaga_initiation_2013}. Water's multitudinous influence throughout planets demonstrates that it is not only the total budget of water that matters for habitability---the partitioning of water between the planet's surface and interior is crucial as well.

Although we may never answer how much water an exoplanet carries in its interior at a given time, we can place some constraints. Solid planetary mantles have thermodynamic limits to water saturation; their total water \textit{capacities} follow deterministically from known and measurable parameters. Further, as upper limits, these are physically meaningful, for they approximate the water budget of a planet's mantle shortly after its formation: primordial mantles may inherit large inventories of water as they crystallise from magma oceans overlain by thick steam atmospheres \citep{tikoo_fate_2017, dorn_hidden_2021, bower_retention_2021, miyazaki_wet_2022, salvador_convective_2023}. In these water-saturated conditions, the solubility of water in hot, newly-solidified mantles provides the initial condition for subsequent water cycling---a particularly important bound for stagnant lid planets lacking an efficient return flux of volatiles to the mantle \citep[e.g.,][]{foley_carbon_2018}. The maximum water capacity of the mantle under a stagnant lid gives a hard upper limit on the ocean mass that it could outgas over geologic time. On young planets, rates of total outgassing also correlate more strongly with the initial water content of the mantle than its temperature \citep{guimond_low_2021}. This could matter, for example, on planets around M-dwarfs, if they must sequester water in their mantles or else lose it early on to space under high stellar XUV radiation \citep{wordsworth_water_2013, luger_extreme_2015, godolt_habitability_2019, fleming_xuv_2020, moore_keeping_2020}; replenishment of any surface oceans may be limited by mantle water capacity.

To make a quantitative estimate of interior water capacities, we consider ``water'' not present as \ce{H2O}, but held as hydroxyl groups in the crystal structure of the nominally-anhydrous minerals (NAMs), such as olivine, that make up the mantle. That is, OH exists only in defects in the mineral, rather than as a stoichiometric component. Each NAM has a thermodynamically-limited water capacity, beyond which defects in the crystal are saturated, and additional water added to the system would be present as a free-water phase. The maximum water storage capacity of a planet's mantle is therefore determined by its mineralogy, and by those minerals' respective water capacities. Although planets have another large reservoir of H in their cores, our study does not count core H towards the interior water capacity because this reservoir is unlikely to participate in subsequent planetary evolution.

A planet's mantle mineralogy, in turn, depends on its elemental composition, pressure, and to a lesser extent, temperature. This elemental composition is set during formation, by a planet's building blocks, which themselves are linked ultimately to the composition of the system's natal molecular cloud \citep{anders_solar-system_1982, thiabaud_elemental_2015, bonsor_host-star_2021}. Therefore exoplanet mantle mineralogies are predicted in tandem with \textit{(i)} measurements of stellar refractory element abundances and \textit{(ii)} models calculating the equilibrium mineralogy for a given composition, pressure, and temperature \citep[e.g.,][]{connolly_geodynamic_2009}, as has been investigated extensively in the literature \citep{dorn_can_2015, dorn_bayesian_2017, dorn_generalized_2017, dorn_new_2019, unterborn_effects_2017, unterborn_nominal_2022, hinkel_starplanet_2018, otegi_impact_2020, spaargaren_influence_2020, spaargaren_plausible_2022, wang_detailed_2022}. In particular, the ratio of the two most abundant refractory lithophile elements, Mg to Si, has a first-order effect on mineralogy \citep[e.g.,][]{dorn_bayesian_2017, unterborn_effects_2017, hinkel_starplanet_2018, wang_enhanced_2019, spaargaren_influence_2020, spaargaren_plausible_2022}. We thus expect Mg/Si to control a planet's mantle water storage capacity as an outcome of stellar nucleosynthesis. Overall mineral proportions will also depend on planet size, as high-pressure mineral phases (e.g., postperovskite) make up a large volume of Earth-sized and larger planets. The wide range in water solubility across NAMs, stable at certain characteristic pressures, means that water capacity will not scale linearly with planet mass.

In pursuit of constraints on the water capacity of Earth's mantle, the last two decades saw several efforts compiling the experimental and theoretical constraints on NAM water solubilities \citep[e.g.,][]{keppler_thermodynamics_2006, ohtani_hydrous_2015, demouchy_distribution_2016, tikoo_fate_2017, dong_constraining_2021, andrault_mantle_2022}. Disagreement may persist for some minerals---high-pressure phases in particular---yet there seems to be a robust shape to the mantle water saturation profile overall. Its most obvious feature is distinct discontinuities where (Mg, Fe)$_2$SiO$_4$ olivine transitions to its high-pressure polymorph wadsleyite, and again where the yet higher-pressure olivine polymorph ringwoodite dissociates to (Mg, Fe)SiO$_3$ perovskite. The olivine-bearing part of the upper mantle tends to have lower water saturation, mostly on the order of hundreds of parts per million by weight (but increasing with pressure). The appearance of wadsleyite marks the mantle transition zone (MTZ); here, water saturations leap to the weight-percent level. The beginning of perovskite stability marks the start of the lower mantle, which most studies find to be drier again---a few experiments have found high perovskite water saturations of $\sim$1 wt\%, however \citep{murakami_water_2002, fu_water_2019}. As a consequence, the MTZ is not just a reservoir for water, but also a key structural feature of the deep water cycle: \textit{(i)} the ringwoodite-perovskite transition acts as a bottleneck for water transport down to the drier, deeper mantle; and, \textit{(ii)} upward flow across the wadsleyite-olivine transition acts as a source of hydrous magma by dehydration melting \citep{bercovici_whole-mantle_2003, andrault_mantle_2022}. A key question in the context of exoplanets is whether the distinct mantle water reservoirs on Earth are ubiquitous among rocky planets.

Important work has already been undertaken investigating mantle water capacities beyond Earth. \citet{shah_internal_2021} calculated the water saturation of idealised mantles with fixed NAM compositions in the Mg-Fe-Si system, with varying planet mass and potential temperature. 
More recently, \citet{dong_water_2022} quantified Mars' mantle water capacity given constraints on its particular bulk composition. We build on these two studies by exploring water capacities over a wide range of possible mantle Mg-Fe-Si-Ca-Al bulk compositions estimated from the Hypatia Catalog of FGKM star element abundances \citep{hinkel_stellar_2014}. 
Because the diversity of mantle water capacities has not been quantified, previous rocky planet water cycling models often use approximate Earth values for their initial conditions and mantle water saturation limits \citep[e.g.,][]{cowan_water_2014, schaefer_persistence_2015, komacek_effect_2016, moore_keeping_2020}. 

This paper builds a method to convert stellar compositions into the interior water capacities of hypothetical rocky planets associated with those stars. In section \ref{sec:methods_comp}, we translate stellar element abundances into planetary oxide bulk compositions. In section \ref{sec:methods_structure}, we self-consistently calculate interior structures and mantle mineralogies from these bulk compositions, as a function of planet mass and potential temperature. Section \ref{sec:methods_sat} then describes how we use mineral water saturation data to estimate interior water capacities. We present our calculated distribution of rocky planet water capacities in section \ref{sec:results}. To end, section \ref{sec:discussion} discusses the implications for whole-planet water cycling and the propensity of waterworlds.

\section{Methods}

\subsection{From stellar abundances to planet bulk composition}\label{sec:methods_comp}

Our hypothetical planets are composed of a solid, pure Fe core and a silicate mantle of \ce{MgO}, \ce{SiO2}, \ce{Al2O3}, \ce{CaO}, and \ce{FeO}\footnote{This implicitly assumes that all oxidised iron exists as \ce{FeO} and not \ce{Fe2O3}, since we do not track mantle oxygen fugacity.} We expect these oxides to make up $\gtrsim$98\% by mass of all rocky exoplanet mantles \citep{putirka_composition_2019}. The planets we model inherit identical bulk metal ratios to their host stars. We assume that oxygen would be abundant enough to bond with Ca, Mg, Al, and Si such that they are fully present as oxides in the mantle---Fe is present in excess of available oxygen and hence produces a metal core to the planet. 

To specify the partitioning of total iron between Fe metal in the core and FeO in the mantle, we define \coreeff~as the molar ratio,
\begin{equation}\label{eq:core_eff}
    \chi^{\rm Fe}_{\rm m} = \frac{n_{\rm m}^{\rm Fe}}{n_{\rm m}^{\rm Fe} + n_{\rm c}^{\rm Fe}},
\end{equation}
where $n_{\rm m}^{\rm Fe}$ and $n_{\rm c}^{\rm Fe}$ are the number of moles of Fe in the mantle and core respectively.

For each planet-hosting FGKM star in the Hypatia Catalog\footnote{https://www.hypatiacatalog.com/} \citep{hinkel_stellar_2014}, including outliers, we find the mantle bulk composition and core mass fraction of hypothetical rocky exoplanets, depending on the star's measured elemental abundances and our choice of \coreeff. Specifically, we extract the number abundance of the rock-forming element X with respect to hydrogen, $n_{\rm X}/n_{\rm H}$, using the mean where multiple measurements exist. Hypatia reports logarithmic abundances, denoted ${\rm [X/H]}$, with respect to a solar normalisation:
\begin{equation}
    {\rm [X/H]} = \log_{10}(n_{\rm X}/n_{\rm H})_* - \log_{10}(n_{\rm X}/n_{\rm H})_{\sun},
\end{equation}
where the first term is the stellar value and the second is the solar value---here, that of \citet{lodders_abundances_2009}, as in \citet{hinkel_starplanet_2018}. The corresponding list of $\left(n_{\rm X}/n_{\rm H}\right)_*$ can then be written as molar ratios of the metals in their respective oxide forms considering their stoichiometry; i.e., $n_{\rm Mg}:n_{\rm Si}:2n_{\rm Al}:n_{\rm Ca}:$ \coreeff$n_{\rm Fe}$. We convert these molar ratios to mass ratios of the respective oxides via molar mass and normalise to 100\%. As such, we make the simplification that these metals are equally refractory: whilst their absolute abundances in the bulk planet may not equal those in the stellar photosphere, we expect the ratios between them to be roughly preserved.  We do not propagate observational errors on stellar abundances, but a thorough discussion of this issue can be found in \citet{hinkel_starplanet_2018} and \citet{hinkel_concise_2022}.

\subsubsection{Core mass fraction and partitioning of Fe}

We assume, in all cases, a pure iron core. The core mass fraction (CMF) is calculated from the mantle FeO content and \coreeff, conserving the bulk planet Fe content. We derive the CMF by the definition of \coreeff (\ref{eq:core_eff}):
\begin{equation}
\begin{split}
    n_{\rm m}^{\rm Fe} &= \frac{f_{\rm FeO} M_p \left(1 - {\rm CMF}\right)}{M_{\rm FeO}} \\
    n_{\rm c}^{\rm Fe} &= \frac{{\rm CMF} \, M_p}{M_{\rm Fe}},
\end{split}
\end{equation}
where $M_{\rm Fe}$ and $M_{\rm FeO}$ are the molar masses of Fe and FeO respectively in ${\rm kg}\,{\rm mol}^{-1}$, $M_p$ (which cancels out) is the mass of the planet in kg, and $f_{\rm FeO}$ is the mantle weight fraction of FeO. This has the analytical solution,
\begin{equation}\label{eq:CMF}
    {\rm CMF} = \frac{M_{\rm Fe} f_{\rm FeO} \left(1 - \chi^{\rm Fe}_{\rm m}\right)}{-M_{\rm Fe} f_{\rm FeO} \chi^{\rm Fe}_{\rm m} + M_{\rm Fe} f_{\rm FeO} +  M_{\rm FeO} \chi^{\rm Fe}_{\rm m}}.
\end{equation}
In this way, an Earth-like FeO mantle fraction \citep[8.05 wt\%;][]{mcdonough_composition_1995} and core mass fraction (0.325) are equivalent to \coreeff$\,= 0.117$; \coreeff$\,= 0$ denotes an Fe-free mantle.



We consider a generous range of \coreeff~from 0 to 0.3. The existence of core-free rocky planets has been proposed before, but these planets may require formation environments out beyond the snow line---where they can accrete very oxidised, water-rich material \citep{elkins-tanton_coreless_2008, kite_atmosphere_2020}. Our range of \coreeff~encompasses mantles more oxidising than Mars', which itself has had its FeO content recently revised from 17--18 wt\% down to $\sim$14 wt\% \citep{khan_geophysical_2022}, giving, in our pure-iron-core model, \coreeff~= 0.236 for a martian core mass fraction of 0.25. 

\subsubsection{Some initial caveats}

In performing these estimates, we are not aiming to predict in detail the composition of a particular planet's mantle, but rather to capture the possible variability in water storage capacity at a population level. Namely, planetary accretion and differentiation may involve several fractionation processes that modify the compositional relationship between star and planet. Being the most refractory out of the elements we consider, Al and Ca show no depletion on Earth with respect to solar abundances---our simplified method will therefore slightly underestimate the proportions of \ce{Al2O3} and \ce{CaO} relative to other oxides. For similar reasons we have excluded Na, a moderately volatile element that would fractionate more strongly with respect to the others, during condensation from the nebular gas and evaporation from molten planets (and is nevertheless an order of magnitude less abundant). 

Meanwhile, Mg, Si, and Fe are expected to have similar depletion ($\sim$14--20\%) between the star and bulk planet \citep{wang_enhanced_2019}---yet we know from Earth that solar Mg/Si ($=1.04$) is only a rough approximation of mantle composition as it fails to predict Earth's measured upper mantle Mg/Si of 1.25; somewhere, partitioning must be occurring \citep[see also section \ref{sec:discussion-sio2}]{ringwood_significance_1989, mcdonough_composition_1995, lodders_abundances_2009}. Nevertheless, despite knowing that in detail stellar Si abundances will not equal those in the planet, the exact processes leading to the incorporation of Si into a growing planet's mantle (or core) are not yet easily and deterministically quantifiable: they involve not just devolatilisation and dust transport in disks \citep{miyazaki_dynamic_2020}, but also the high-pressure core-mantle partitioning of Si \citep{fischer_high_2015}.

Further to mantle compositions, the cores of rocky planets may not be pure solid Fe. Cores could contain lighter elements, or be partially or entirely molten. Both phenomena would affect the core radius at a given CMF, and accordingly the pressure and temperature gradients of the mantle \citep{unterborn_scaling_2016}. Although these particular effects of changing core radius are inconsequential to our water capacity results, the principles are discussed in section \ref{sec:results_fe}. Note that if one of these light elements is Si, then the mantle mineralogy changes (section \ref{sec:discussion-sio2}). If one of them is H, then the core could in this way contribute to the total ``water'' capacity of the planet \citep{shah_internal_2021}. Nevertheless, the solubility of water in mantle minerals is independent from that in the core at water saturation. Even more importantly for planetary evolution, the core H reservoir is likely sealed from the surface following its formation, whereas mantle water may be released through volcanism.

\subsection{Interior structure and mineralogy}\label{sec:methods_structure}

To compute equilibrium mineral phase abundances given a bulk oxide composition, we use the Gibbs free energy minimisation code {\tt Perple\_x} \citep{connolly_geodynamic_2009}. This code, with some extrapolation of thermodynamic properties, has previously been used to investigate the silicate mantles of planets up to 10 $M_\oplus$ \citep{dorn_can_2015, dorn_generalized_2017, dorn_new_2019, unterborn_inward_2018, hinkel_starplanet_2018, unterborn_pressure_2019, otegi_impact_2020, wang_detailed_2022}. At the current time we are not able to consider the effect of water on sub-solidus phase boundaries, although experiments on water-present, Earth-like mantle compositions suggest a limited influence of water incorporation on NAM stability \citep[shifting the depth of phase boundaries by at most a few GPa;][]{litasov_influence_2006}.

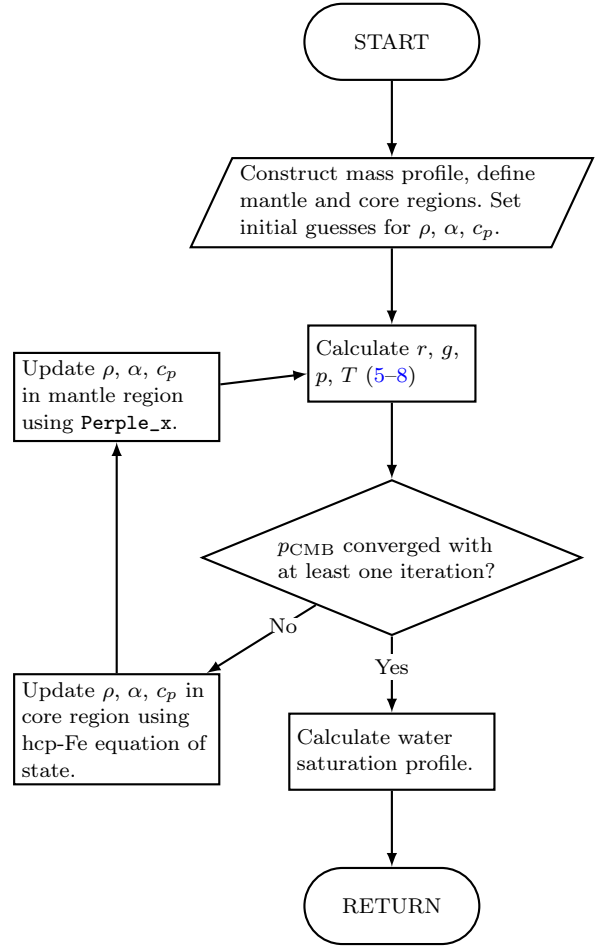
\begin{figure}
\caption{Interior model iterative method. See text for symbol definitions and further details. \label{fig:flowchart}}

\begin{tikzpicture}[font=\small,thick]

\node[draw,
    rounded rectangle,
    minimum width=2.5cm,
    minimum height=1cm] (block1) {START};
    
    \node[draw, text width=4cm,
    trapezium, 
    trapezium left angle = 65,
    trapezium right angle = 115,
    trapezium stretches,
    below=of block1,
    minimum width=3.5cm,
    minimum height=1cm
] (block2) { Construct mass profile, define mantle and core regions. Set initial guesses for $\rho$, $\alpha$, $c_p$. };

\node[draw, text width=2cm,
    below=of block2,
    minimum width=1cm,
    minimum height=1cm
] (block3) { Calculate $r$, $g$, $p$, $T$ (\ref{eq:radius_prof}--\ref{eq:temperature_prof})};

\node[draw,  text width=3cm,
    diamond, aspect=2,
    below=of block3,
    minimum width=5cm,
    inner sep=0] (block4) { $p_{\rm CMB}$ converged with at least one iteration?};

        \node[draw,  text width=2.5cm,
    below left=of block4,
    minimum width=2.5cm,
    minimum height=1cm] (block5) { Update $\rho$, $\alpha$, $c_p$ in core region using hcp-Fe equation of state.};

\node[draw, text width=2.5cm,
    above left=of block4,
    minimum width=2.5cm,
    minimum height=1cm] (block6) { Update $\rho$, $\alpha$, $c_p$ in mantle region using {\tt Perple\_x}.};
    
        \node[draw,  text width=2.5cm,
    below=of block4,
    minimum width=2.5cm,
    minimum height=1cm,] (block7) { Calculate water saturation profile.};

\node[draw,
    rounded rectangle,
    below=of block7,
    minimum width=2.5cm,
    minimum height=1cm,] (block8) { RETURN};
    
    \draw[-latex] (block1) edge (block2)
    (block2) edge (block3)
    (block3) edge (block4)
    (block7) edge (block8)
    (block5) edge (block6)
    (block6) edge (block3);
    
\draw[-latex] (block4) edge node[pos=0.4,fill=white,inner sep=2pt]{Yes}(block7)
    (block4) edge node[pos=0.3,fill=white,inner sep=0]{No} (block5);

\end{tikzpicture}
\end{figure}

Phase abundances, being a function of pressure and temperature, must be calculated along an adiabat from the surface to the core-mantle boundary. However, the adiabatic gradient depends in turn on the phases present. Therefore our model self-consistently finds the mantle pressure $p$, temperature $T$, and phase abundance profiles using an iterative method summarised in Figure \ref{fig:flowchart}.

Given an input planet mass, we subdivide a fixed mass profile, $m$ in kg, into $N$ equal shells from 0 to $M_p$, where $300 < N < 1600$ depending on $M_p$. At each iteration, we build the profiles of radius, $r$ in m, and gravity, $g$ in ${\rm m}\,{\rm s}^{-2}$, upwards from planet-centre boundary conditions of $r_0 = g_0 = 0$, 
\begin{equation}\label{eq:radius_prof}
r_n = \left(\frac{3\Delta m }{4 \pi \rho_n} + r_{n - 1}^3\right)^{\frac{1}{3}},
\end{equation}
\begin{equation}\label{eq:gravity_prof}
g_n = g_{n - 1} \left(\frac{r_{n - 1}}{r_n}\right)^2 + \frac{4 \pi G}{3} \rho_n \left(\frac{r_n^3 - r_{n - 1}^3}{r_n^2}\right)
\end{equation}
where $\Delta m = m_n - m_{n - 1}$ in kg (constant), $G$ is the gravitational constant in ${\rm m}^3 \, {\rm kg}^{-1}\,{\rm s}^{-2}$, and $\rho_n$ is the density of layer $n$ in ${\rm kg} \, {\rm m}^{-3}$.

We build the adiabatic $p$--$T$ profiles downwards from their near-surface boundary conditions: $p_{\rm sfc} = 1000$~bar and a $T_{\rm sfc}$ equal to the input potential temperature, $T_p$, in K:
\begin{equation}\label{eq:pressure_prof}
p_{n} = p_{n + 1} + \Delta r \, g_{n + 1} \rho_{n + 1},
\end{equation}
\begin{equation}\label{eq:temperature_prof}
T_{n} = T_{n + 1} + \Delta r \, \frac{\alpha_{n}g_{n}}{c_p,_{n}} T_{n + 1},
\end{equation}
where $\Delta r = r_n - r_{n - 1}$ in m, $c_p$ is the specific heat capacity in ${\rm J}\,{\rm kg}^{-1}\, {\rm K}^{-1}$ and $\alpha$ is the thermal expansivity in ${\rm K}^{-1}$; initial guesses for the mantle and core layers respectively are $1300\,{\rm J}\,{\rm kg}^{-1}\, {\rm K}^{-1}$ and $800\,{\rm J}\,{\rm kg}^{-1}\, {\rm K}^{-1}$ for $c_p$, and $2.5 \times 10^{-5}\,{\rm K}^{-1}$ and $1 \times 10^{-5}\,{\rm K}^{-1}$ for $\alpha$. Initial guesses for $\rho$ through the mantle and core layers are the average values from the parameterised interior structure of \citet{noack_parameterisations_2020}. For simplicity, we ignore any temperature jumps due to phase transitions or core heat transfer. 

We identify the core-mantle boundary pressure, $p_{\rm CMB}$, as the pressure at the layer $n$ where the cumulative mass reaches the pure-Fe core mass from (\ref{eq:CMF}): $\sum_n m_n \ge {\rm CMF} \cdot M_p$.

At subsequent iterations, we run {\tt Perple\_x} to find the mineral phase mass fractions given both the mantle $p$--$T$ profile from the previous iteration, and the mantle bulk oxide composition from section \ref{sec:methods_comp}. The {\tt Perple\_x} output consists of new $\rho$, $\alpha$, and $c_p$ profiles consistent with the calculated phase fractions. Our {\tt Perple\_x} implementation uses the most recent mineral solution models and equations of state from \citet{stixrude_thermal_2022}. The 22-phase coverage of the \citet{stixrude_thermal_2022} database should be mostly complete for exoplanet mantles because wholly exotic bulk compositions are not expected \citep{putirka_composition_2019}. Note that other thermodynamic databases could result in different phase fractions; the \citet{stixrude_thermodynamics_2011} database, for example, produces a lower mantle with relatively less perovskite and more ferropericlase at an Earth-like bulk composition.

Above $200\,{\rm GPa}$, the thermodynamic data available to predict mantle mineralogy becomes very limited. Therefore, to avoid phase stability errors in the lower mantles of our calculated planets with $p_{\rm CMB} \ge 200\,{\rm GPa}$, we extrapolate constant mineralogies from $200\,{\rm GPa}$ to $p_{\rm CMB}$, although in this case we still calculate thermodynamic parameters using the same equations of state. At $M_p \gtrsim 3 M_\oplus$, the maximum mantle pressures found in our model start to surpass the theoretical pressures where Mg-postperovskite recombines with MgO or SiO$_2$ \citep{umemoto_phase_2017}. Whilst we still estimate a mantle water capacity for these higher-mass planets, we note that their lower mantle mineralogies will be much more uncertain. 

At pressures above $p_{\rm CMB}$, we calculate $\rho$, $\alpha$, and $c_p$ using a Holzapfel equation of state for pure hcp iron with parameters from \citet{bouchet_ab_2013} and \citet{hakim_new_2018}. 

Given the updated $\rho$, $\alpha$, and $c_p$, we re-calculate $r$, $g$, $p$, and $T$ (\ref{eq:radius_prof}--\ref{eq:temperature_prof}) to find the new adiabat, and so on. We stop the iteration when $p_{\rm CMB}$ has converged to within 0.01\%, which normally takes less than 10 iterations. This precision meets our needs, being, ultimately, to obtain $T$, $p$, $m$, and phase abundance profiles of a planet's mantle.

Our interior model is similar to ExoPlex \citep{unterborn_inward_2018}, and results in the same pressure, temperature, density, and phase gradients when using the same thermodynamic database.

\subsection{Mineral water solubility parameterisations}\label{sec:methods_sat}

\begin{table*}
\centering
\caption{Water saturation model parameter values, either parameterised via (\ref{eq:c_h2o}) and (\ref{eq:D_ji}), or given mineral-mineral partitioning coefficients ($D$) or water concentrations ($c_{\rm H_2O}$). See text for details. \label{tab:saturation_params}}
\footnotesize
\begin{tabular}{@{} l r r r r    r r p{5cm}}
\toprule
 & \multicolumn{4}{c}{Parameters in (\ref{eq:c_h2o})} & \multicolumn{2}{c}{Data range} \\ \cmidrule(lr){2-5} \cmidrule(lr){6-7}
 
\multirow{2}{*}{Phase} & \multicolumn{1}{c}{$A$} & \multicolumn{1}{c}{\multirow{2}{*}{$n$}} & \multicolumn{1}{c}{$\Delta H^{\mathrm{1\,bar}}$} & \multicolumn{1}{c}{$\Delta V^{\mathrm{solid}}$} & \multicolumn{1}{c}{$T$} & \multicolumn{1}{c}{$p$} & \multirow{2}{*}{Ref.} \\

 & \multicolumn{1}{c}{$({\rm ppm\,bar^{-n}})$} & & \multicolumn{1}{c}{$({\rm kJ\,mol}^{-1})$} & \multicolumn{1}{c}{$({\rm cm^3\,mol}^{-1})$} & \multicolumn{1}{c}{$({\rm K})$} & \multicolumn{1}{c}{$({\rm GPa})$} & \\
 
\hline
\noalign{\vskip 1mm}
\multicolumn{8}{c}{$\bm{c}\mathbf{_{H_2O}}$ \textbf{calculated directly}} \\
\noalign{\vskip 1mm}

ol &  $4.6838\times 10^{-2}$ & 0.33235 & $-40.787$ & 0 & 1273--2273 & 1.8--14.5 & \citet{dong_constraining_2021}  \\  
wad &  $4.8295\times 10^{-4}$ & 0 & $-114.230$ & 0 & 1673--2273 & 13.5--20.0 & \citet{dong_constraining_2021}  \\  
ring & $1.0224\times 10^{-3}$ & 0 & $-101.482$ & 0 & 1673--2073 & 17--23 & \citet{dong_constraining_2021}  \\  
coes & \multicolumn{4}{c}{$T, p$-dependent $c_{\rm H_2O}$ (see text)} & 1073--1573 & 5.0--9.1 & \citet{yan_water_2021}  \\
stv & \multicolumn{4}{c}{$T, p$-dependent $c_{\rm H_2O}$ (see text)} & 1000--2000 & 0--60 & \citet{panero_hydrogen_2004} \\ 

\noalign{\vskip 1mm}
\hline
\noalign{\vskip 1mm}
\multicolumn{8}{c}{$\bm{c}\mathbf{_{H_2O}}$ \textbf{calculated using a parameterised partitioning coefficient with ol}} \\
\noalign{\vskip 1mm}

opx (en)$^{\ast\dagger\ddagger}$ & $1.354\times 10^{-2}$ & 1.0 & $-4.563$ & 12.1 & 1373 & 2--10 & \citet{rauch_water_2002} \\
opx (Al-)$^{\ast\dagger\ddagger}$ & $4.2\times 10^{-2}$ & 0.5 & $-79.685$ & 11.3 & 1073--1373 & 1.5--3.5 & \citet{mierdel_water_2007} \\
cpx$^{\dagger}$ & 7.144 & 0.5 & 0 & 8.019 & 873--973 & 2--10 & \citet{bromiley_experimental_2004} \\
ol$^{\dagger\mathsection}$ & $6.6\times 10^{-3}$ & 1.0 & 0 & 10.6 & 1273--1373 & 2.5--19.5 & \citet{kohlstedt_solubility_1996} \\

\noalign{\vskip 1mm}
\hline
\toprule
\noalign{\vskip 1mm}
 & \multicolumn{1}{c}{Phases} & \multicolumn{2}{c}{$D$} & \multicolumn{4}{l}{$\bm{c}\mathbf{_{H_2O}}$ \textbf{calculated by partitioning otherwise}} \\
 \cmidrule(lr){2-4}
\noalign{\vskip 1mm}

ppv &  \multicolumn{1}{l}{pv-ppv} &  \multicolumn{2}{c}{$\sim$2--4 (see text)} & & 2000--2900 & $\sim$120 & \citet{townsend_water_2016} \\
hcpx & \multicolumn{1}{l}{ol-hcpx} & \multicolumn{2}{c}{0.7} & & 1473--1673 & 8--11 & \citet{withers_h2o_2007}  \\
gt & \multicolumn{1}{l}{ol-gt} & \multicolumn{2}{c}{1}  & & 1373--1900 & 5--24 & \citet{mookherjee_solubility_2010, ardia_h2o_2012, ferot_water_2012, novella_distribution_2014, liu_bridgmanite_2021} \\

\noalign{\vskip 1mm}
\toprule
\noalign{\vskip 1mm}
 & \multicolumn{2}{c}{$c_{\rm H_2O}\,({\rm ppm\,wt})$} & \multicolumn{5}{l}{$\qquad\qquad\qquad\qquad\bm{c}\mathbf{_{H_2O}}$ \textbf{assumed constant}} \\ \cmidrule(lr){2-3}
\noalign{\vskip 1mm}

pv &30 & & & & 1700--1900 & 24--26 & \citet{liu_bridgmanite_2021} \\
fp & 10 & && & 1673--2273 & 25 & \citet{litasov_influence_2010} \\
dvm & 5000 && && 1400--2200 & 19--120 & \citet{chen_possible_2020} \\
seif & 1000 & & & & 1380--3300 & 44--152 & \citet{lin_hydrous_2022} \\

\noalign{\vskip 1mm}
\bottomrule
\noalign{\vskip 1mm}
\multicolumn{8}{l}{$^\ast$\scriptsize{$c_{\rm opx}$ is the 1:1 sum of pure enstatite (en) and aluminous opx.}}\\
\multicolumn{8}{l}{$^\dagger$\scriptsize{$f_{\rm H_2O}$ is calculated following \citet{pitzer_equations_1994} for consistency with the original fit.}}\\
\multicolumn{8}{p{\textwidth}}{$^\ddagger$\scriptsize{A factor of 3 is applied to $c_{\rm ol}$ from \citet{kohlstedt_solubility_1996}, to convert the calibration of \citet{paterson_determination_1982}, which is used in their fit, to the calibration of \citet{bell_quantitative_1995}, which is used in the fit in the reference. Note the \citet{paterson_determination_1982} calibration is known to be an underestimate \citep{keppler_thermodynamics_2006, bolfan-casanova_examination_2018}.}}\\
\multicolumn{8}{l}{$^\mathsection$\scriptsize{Used only with (\ref{eq:c_h2o}) to calculate the partitioning of water between ol and the phases in this section.}}\\
\end{tabular}
\end{table*}

The saturation water content of a NAM is interpreted as the OH concentration of the mineral in equilibrium with an aqueous fluid, or---at $p$ and $T$ beyond the critical point---with a hydrous melt, if the phase assemblage is invariant \citep{keppler_thermodynamics_2006}. From the Gibbs free energy of the solution reaction one obtains:
\begin{equation}
    c_{\rm H_2O} = A \left(f_{\rm H_2O}\right)^n \exp\left(-\frac{\Delta H^{\mathrm{1\,bar}} + \Delta V^{\mathrm{solid}} p}{R_b T}\right),
\label{eq:c_h2o}
\end{equation}
with $c_{\rm H_2O}$ in ppm by weight, $A$ in ${\rm ppm}$, the oxygen fugacity, $f_{\rm H_2O}$, $n$ relating to the OH substitution mechanism, the reaction enthalpy, $\Delta H^{\mathrm{1\,bar}}$ in ${\rm J\,mol}^{-1}$, the volume change of the solids, $\Delta V^{\mathrm{solid}}$, in ${\rm m^3\,mol}^{-1}$, $p$ in Pa, the gas constant $R_b = 8.314\; {\rm J\, mol^{-1}\,K^{-1}}$, and $T$ in K.

At water saturation, the presence or absence of other phases will not affect the equilibrium water content of any other phase: water is available to fill all minerals to their maximum capacity. Thus, provided the water saturation content of one phase is known, the water content of other phases at saturation can be determined from the known partitioning of water between them. The partition coefficient of water between phases $i$ and $j$ is given by:
\begin{equation}
    D^j_i = \frac{c_j}{c_i},
    \label{eq:D_ji}
\end{equation}
where $c_j$ and $c_i$ are the water concentrations at saturation of the phases; e.g., from (\ref{eq:c_h2o}).

The maximum water mass fraction of a layer $k$ is the linear combination of the water saturation concentrations across all phases present:
\begin{equation}
    c_{{\rm tot}, k} = \sum_i X_i c_i,
\end{equation}
where $X_i$ is the mass fraction of phase $i$ given by {\tt Perple\_x} (section \ref{sec:methods_structure}). Similarly, the mass of water present in a planetary mantle at water saturation in kg is:
\begin{equation}\label{eq:water_mass}
    w = \sum_k c_{{\rm tot}, k} m_k,
\end{equation}
where $m_k$ is the mass of layer $k$ in kg, calculated as described in section \ref{sec:methods_structure}. We are interested in the maximum water mass of the whole mantle, $w_{\rm m}$, in addition to the maximum water mass in just the upper mantle, $w_{\rm um}$. Here the upper mantle (which includes the MTZ) integrates all layers at lower pressure than the shallowest occurrence of perovskite. The rest of this section details how we calculate the water solubility of each mineral and is summarised in Table \ref{tab:saturation_params}.

\subsubsection{Water solubility of olivine, wadsleyite, and ringwoodite}

For the common upper mantle mineral olivine (ol) and its high-pressure polymorphs, we follow \citet{dong_constraining_2021}, who fit the available data to an equation of similar form to (\ref{eq:c_h2o}). Table \ref{tab:saturation_params} converts their fit to the equivalent parameters in (\ref{eq:c_h2o}). Here, $f_{\rm H_2O}$ is calculated with the \citet{frost_experimental_1997} parameterisation.

The water solubility of olivine has been shown to increase with its FeO content, at least at 3--$6\,{\rm GPa}$ and $1473\,{\rm K}$ \citep{withers_effect_2011}. We do not directly parameterise water solubility in terms of olivine Fe content because (i) the measured variability in $c_{\rm ol}$ is within the spread appearing in other experimental data that do not necessarily control for FeO content, and (ii) we cannot necessarily extrapolate the effect of olivine FeO content on water solubility to higher pressures. The bulk FeO content of the mantle itself (already a free parameter via \coreeff) does nonetheless affect water capacity through its effect on modal mineralogy.

\subsubsection{Water solubility of upper mantle and transition zone minerals}

The water capacities of other minerals stable in the upper mantle are calculated via their partitioning of water with olivine. In the case of some pyroxenes, experimental data has produced a thermodynamically-formulated parameterisation of $c_{\rm H_2O}$ in terms of $T$, $p$, and $\log f_{\rm H_2O}$, such that the water saturations of both phases can be calculated using (\ref{eq:c_h2o}) and then the partitioning calculated using (\ref{eq:D_ji}). In other cases, data only support the use of a constant $D$ value. Nonetheless, when a parameterisation is available, following \citet{dong_constraining_2021} we use the fit parameters for (\ref{eq:c_h2o}) from \citet{kohlstedt_solubility_1996} to calculate the ``thermodynamic'' $D^{\rm ol}_{\rm opx}$, for example---then we divide $c_{\rm ol}$ from \citet{dong_constraining_2021} by this $D^{\rm ol}_{\rm opx}$ to find $c_{\rm opx}$. 

Note that some bulk mantle compositions, in particular those with high Si/Mg, will not stabilise ol. In these cases, we still calculate a fictive ol water saturation, and find the water partition coefficient with respect to the fictive ol using (\ref{eq:D_ji}). However, if the phase in question has a deeper stability field than ol, we avoid extrapolating $c_{\rm ol}$ to egregious $p$--$T$ when using (\ref{eq:D_ji}), instead capping $c_{\rm ol}$ at its 14.5 GPa value.

\textbf{Orthopyroxene (opx)}---the water saturation of opx changes in the presence of Al-bearing species because in this situation H will occupy an Al defect \citep{keppler_thermodynamics_2006}. Thus, for the numerator of (\ref{eq:D_ji}) we use the evenly-weighted sum of $c_{\rm opx}$ parameterisations for aluminous opx from \citet{mierdel_water_2007} and pure enstatite from \citet{mierdel_temperature_2004}. We find that calculating $c_{\rm opx}$ by partitioning with respect to \citet{kohlstedt_solubility_1996} accords better with $D^{\rm ol}_{\rm opx}$ from more recent experiments \citep{ferot_water_2012, demouchy_subsolidus_2017}, compared to the raw $c_{\rm opx}$ from these studies, whilst providing an explicit $p$--$T$ dependence. Along a 1600 K adiabat this results in $D^{\rm ol}_{\rm opx}$ increasing from 0.2 at $1\,{\rm GPa}$ to 3.1 at $10.7\,{\rm GPa}$.

\textbf{Clinopyroxene (cpx)}---\citet{bromiley_experimental_2004} provide the only temperature-dependent parameterisation for cpx water partitioning in the form of (\ref{eq:c_h2o}), in this case for the cpx endmember NaAlSi$_2$O$_6$ jadeite (we assume a similar water saturation for other cpx compositions). The partitioning $D^{\rm ol}_{\rm cpx}$ resulting from this parameterisation generally agrees with more recent experiments \citep{kovacs_experimental_2012, demouchy_subsolidus_2017, padron-navarta_subsolidus_2017}.

\textbf{High-pressure clinopyroxene (hcpx)}---experiments on hcpx water solubility are summarised in \citet{withers_h2o_2007}. We adopt their average $D^{\rm ol}_{\rm hcpx} = 0.7$ in the ol-hcpx co-stability field of 8--$11\,{\rm GPa}$. 

\textbf{Garnet (gt)}---garnets cover a wide range of compositions; the solubility of a particular garnet has a strong dependence on its Mg content and the oxygen fugacity \citep[e.g.,][]{mookherjee_solubility_2010, zhang_effects_2022}. For example, the pure pyrope endmember of garnet is significantly more water-poor than its majorite or grossular endmembers. At present there is insufficient data on garnet solubility to justify accounting for all possible factors. Thus, following \citet{demouchy_distribution_2016} and \citet{andrault_mantle_2022}, we assume a partition coefficient $D^{\rm ol}_{\rm gt} = 1$, which is a simplification that will nonetheless capture the general trends.

\textbf{Minor phases}---in addition to the above, akimotoite and spinel may also occur in small proportions and over limited regions of $p$--$T$ space (\textless 10~ppm of the mantle by mass). For spinel, which is known to be incorporate very little water, we assume a fixed water solubility of 1~ppm, and for akimotoite we use $D^{\rm rw}_{\rm aki} = 21$ \citep{keppler_thermodynamics_2006}.

\subsubsection{Water solubility of lower mantle minerals}\label{sec:sat_lm}

\hspace{\parindent} \textbf{Perovskite (pv)}---experiments on pv water saturation have been notoriously hard to reconcile. In particular, synthesised crystals must be large enough to ascertain that Fourier-transform infrared spectroscopy (FTIR) measurements are not mistakenly targeting hydrous inclusions, rather than H incorporated in defects. Recently, \citet{liu_bridgmanite_2021} performed an FTIR measurement on a large single crystal of the Mg-endmember bridgmanite at 1900 K. We use their upper limit of $c_{\rm pv} = 30$~ppm. This would imply $D^{\rm ring}_{\rm pv} \approx 21$, which is within the high end of the uncertainty of measurements by \citet{inoue_water_2010}. 

\textbf{Postperovskite (ppv)}---the only information on ppv water saturation comes from theoretical predictions of its partitioning with bridgmanite by \citet{townsend_water_2016}, who find that $D^{\rm pv}_{\rm ppv}$ depends on Al content. We assume aluminous endmember phases are present---i.e., \ce{Al2O3}-(post)perovskite and \ce{FeAlO3}-perovskite---and hence extract the value of $D^{\rm pv}_{\rm ppv}$ from figure 4 of \citet{townsend_water_2016} at our calculated pv-ppv transition temperatures.

\textbf{Ferropericlase (fp)}---(Mg, Fe)O, known as ferropericlase or magnesiow\"ustite, is generally accepted to be dry, having on the order of tens of ppm by weight of water \citep{bolfan-casanova_pressure_2002, litasov_influence_2010}. Like \citet{dong_constraining_2021}, we assume $c_{\rm fp} = 10\,{\rm ppm}$.

\textbf{Davemaoite (dvm)}---some experimental evidence has suggested that \ce{CaSiO3} (also known as calcium perovskite) is an important water carrier in Earth's lower mantle \citep{chen_possible_2020}. Whilst wet dvm is also supported by density functional theory calculations \citep{shim_water_2022}, the FTIR measurements are possibly contaminated from hydrous inclusions \citep{chen_possible_2020}. We assume $c_{\rm dvm} = 0.5$~wt\%, at the low end of the wet scenario. Recently, \citet{ko_calcium_2022} have proposed that dvm may not in fact exist as a separate phase, and all \ce{CaSiO3} dissolves into pv instead. At face value, this disappearance of dvm would imply lower overall mantle water capacities compared to what we model here because generally $c_{\rm pv} < c_{\rm dvm}$.

\textbf{FeAlO$_3$ perovskite (fapv)}---there is no data on the water saturation of the minor, theoretical phase fapv. We assume $c_{\rm fapv} = c_{\rm ppv}$.

\subsubsection{Hydrous phases}

We do not consider additional water capacity due to hydrous minerals; i.e., minerals where water is incorporated stoichiometrically into the crystal structure. Whilst they have been proposed to be an important carrier of water to Earth's lower mantle, hydrous minerals (with possible exceptions; e.g., $\delta$-AlOOH) are generally only stable up to $\sim60\,{\rm GPa}$ and at the cold temperatures of a subducting slab \citep{ohtani_hydrous_2015}. Therefore they may be less important for the fiducial exoplanet scenario, where the planet is in a non-plate tectonic geodynamic regime. This is especially the case for larger, hotter planets, the interiors of which will be further from hydrous mineral stability fields. Hydrous minerals are also not included in the \citet{stixrude_thermal_2022} thermodynamic data, so we would need to employ a different thermodynamic database and set of mineral models to predict them. We note though that over time, hydration of the crust through chemical weathering may become an important planetary water reservoir \citep[e.g.,][]{scheller_long-term_2021}.

\subsubsection{Water solubility of SiO$_2$ polymorphs}\label{sec:methods_sio2}

Notwithstanding the last paragraph, free silica phases are NAMs which may undergo hydration at lower mantle $p$, $T$ and beyond, incorporating several weight percent \ce{H2O} \citep{nisr_large_2020}. However, even anhydrous \ce{SiO2} phases may be important water carriers \citep{panero_hydrogen_2004, litasov_high_2007}. Whilst silica phases are generally not abundant in Earth's Mg-rich lower mantle, the most Si-rich stellar compositions are associated with planetary lower mantles of up to $\sim$30\% silica by volume. 

\textbf{Quartz (qtz)}---we assume qtz is dry and set $c_{\rm qtz} = 1\,{\rm ppm}$. This will not significantly impact our results, as when qtz does appear it is only stable to $\sim3\,{\rm GPa}$, a small fraction of the mass of planets in our study.

\textbf{Coesite (coes)}---for the next-highest-pressure silica polymorph we use the temperature-dependent parameterisation of water solubility from \citet{yan_water_2021}: $c_{\rm coes} = -105 + 5.2 p + 0.112 T$, with $c$ in ppm, $p$ in GPa, and $T$ in K.

\textbf{Stishovite (stv)}---the water capacity of anhydrous stv appears to be controversial. Whilst some studies have found low water capacities of pure stv closer to 300--$400\,{\rm ppm}$ \citep{liu_bridgmanite_2021}, the solubility of water in stv may be much higher in the presence of Al \citep{litasov_high_2007}. For anhydrous aluminous stv we extract the theoretical prediction from figure 4 of \citet{panero_hydrogen_2004}. This results in $c_{\rm stv}$ increasing with $p$ and $T$ to up to $3\,{\rm wt\%}$ at $60\,{\rm GPa}$, which happens to be at the low end of water saturations for hydrous stv \citep{lin_evidence_2020}. 

\textbf{Seifertite (seif)}---there is no experimental data available on the water capacity of anhydrous seif, which is only stable at very high lower-mantle pressures above $120\,{\rm GPa}$. To be consistent with our treatment of stv, we use a fixed value near the minimum water saturation of hydrous seif \citep{lin_hydrous_2022}, at $c_{\rm seif} = 1\,{\rm wt\%}$

\begin{figure*}
         \centering
         \includegraphics[width=0.99\textwidth]{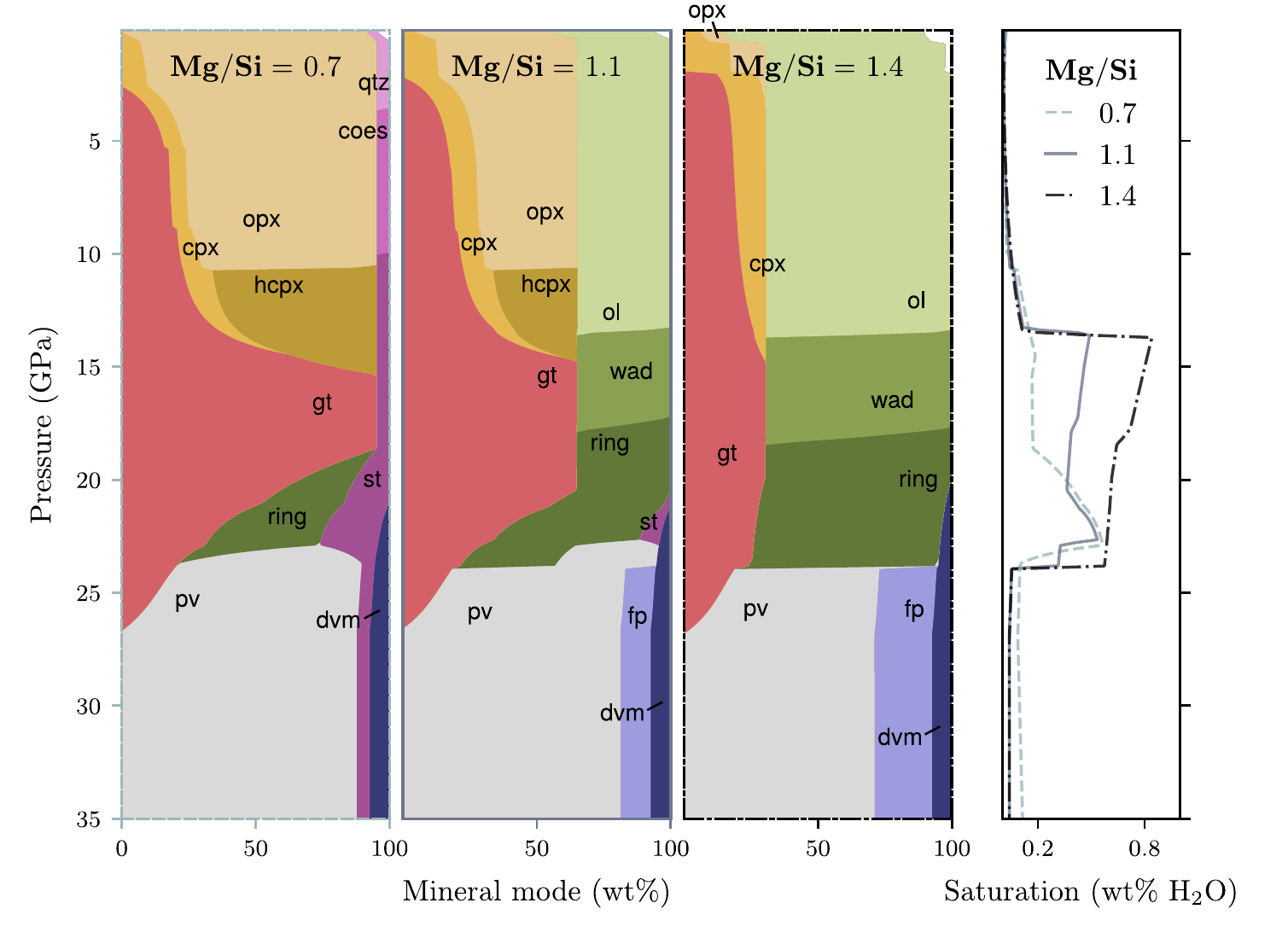}
        \caption{Profiles of mineral mode \textit{(left three panels)} and water saturation content \textit{(right panel)}, for planets with an otherwise Earth-like composition but a molar Mg/Si ratio adjusted to 1.4 \textit{(left mode panel; dashed profile)}, 1.1 \textit{(centre mode panel; solid profile)}, or 0.7 \textit{(right mode panel; dashed-dotted profile)}, respectively representing the 2nd, 50th, and 98th percentiles of exoplanet host stars in the Hypatia Catalog. Adjusted compositions are calculated conserving the total mass of MgO + SiO$_2$. Note that much of the lower mantle pressure range is not shown. Mineralogies and water saturations are for a potential temperature of 1600 K, mantle:bulk Fe of 0.113, and planet mass of $1\,M_\oplus$. Potential temperatures do not have a significant effect on the equilibrium mineralogy. For Mg/Si from 0.7 to 1.1 to 1.4, upper mantle water capacities are 0.9, 1.3 and 2.0 Earth oceans, and total mantle water capacities (i.e., including the lower mantle) are 4.7, 2.4, and 3.1 Earth oceans. Our nominal bulk silicate Earth composition is 45.0 wt\% SiO$_2$, 37.8 wt\% MgO, 8.05 wt\% FeO, 4.45 wt\% Al$_2$O$_3$, and 3.55 wt\% CaO \citep{mcdonough_composition_1995}. Abbreviated phases are garnet (gt), clinopyroxene (cpx), orthopyroxene (opx), high-pressure clinopyroxene (hcpx),  olivine (ol), wadsleyite (wad), ringwoodite (ring), perovskite (pv), quartz (qtz), coesite (coes), stishovite (stv), ferropericlase (fp), and davemaoite (dvm).}
        \label{fig:mgsi_modality}
\end{figure*}

\begin{figure}
    \centering
    \includegraphics[width=0.4\textwidth]{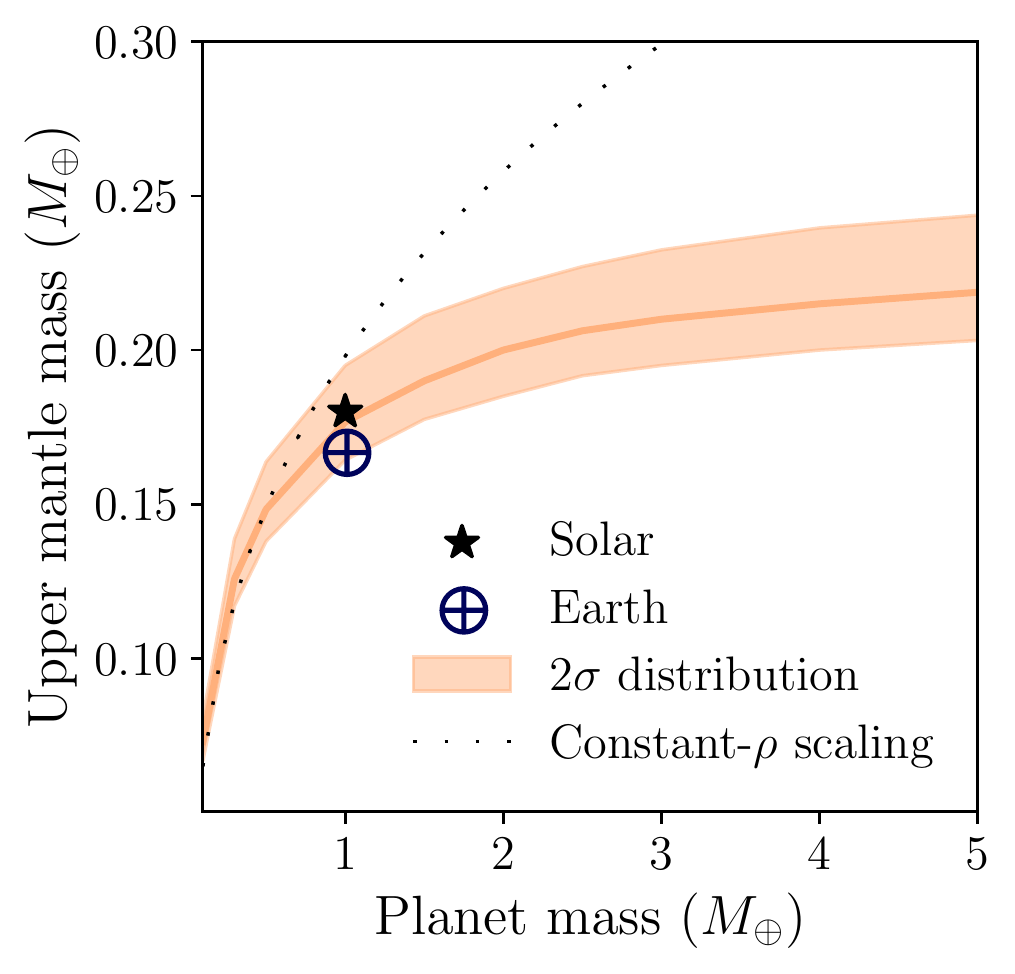}
    \caption{Scaling of the upper mantle mass with the total planet mass. The upper mantle is defined as the shell bounded by the surface and the first occurrence of perovskite. The solid orange line shows the median upper mantle mass, and the swath spans the 2$\sigma$ distribution across compositions of planet-hosting stars from the Hypatia Catalog ($N = 1285$). Results are shown for a potential temperature of 1600 K and a mantle:bulk Fe of 0.113. Note that an FeO-free mantle would result in 5--10\% less-massive upper mantles on average, compared to those shown here; potential temperature has no effect. The $\oplus$ symbol shows the modelled result for an Earth-like bulk composition \citep{mcdonough_composition_1995}, which at $1.04 \times 10^{24}\,{\rm kg}$ is close to the observed value of $1.06 \times 10^{24}\,{\rm kg}$ \citep{nolet_earths_2011}. The star shows an Earth-mass planet with solar bulk composition; i.e., a lower Mg/Si than observed \citep{lodders_abundances_2009}. The dotted line represents a simplified scaling assuming a constant mantle density of $4500\,{\rm kg\,m^{-3}}$ and an Earth-like bulk density of $5510\,{\rm kg\,m^{-3}}$, which overestimates the upper mantle mass at large $M_p$ because compression dictates that massive planets become denser and contain proportionately more mass at depth.}
    \label{fig:mass_um}
\end{figure}

\begin{figure*}
\centering
  \includegraphics[width=0.9\linewidth]{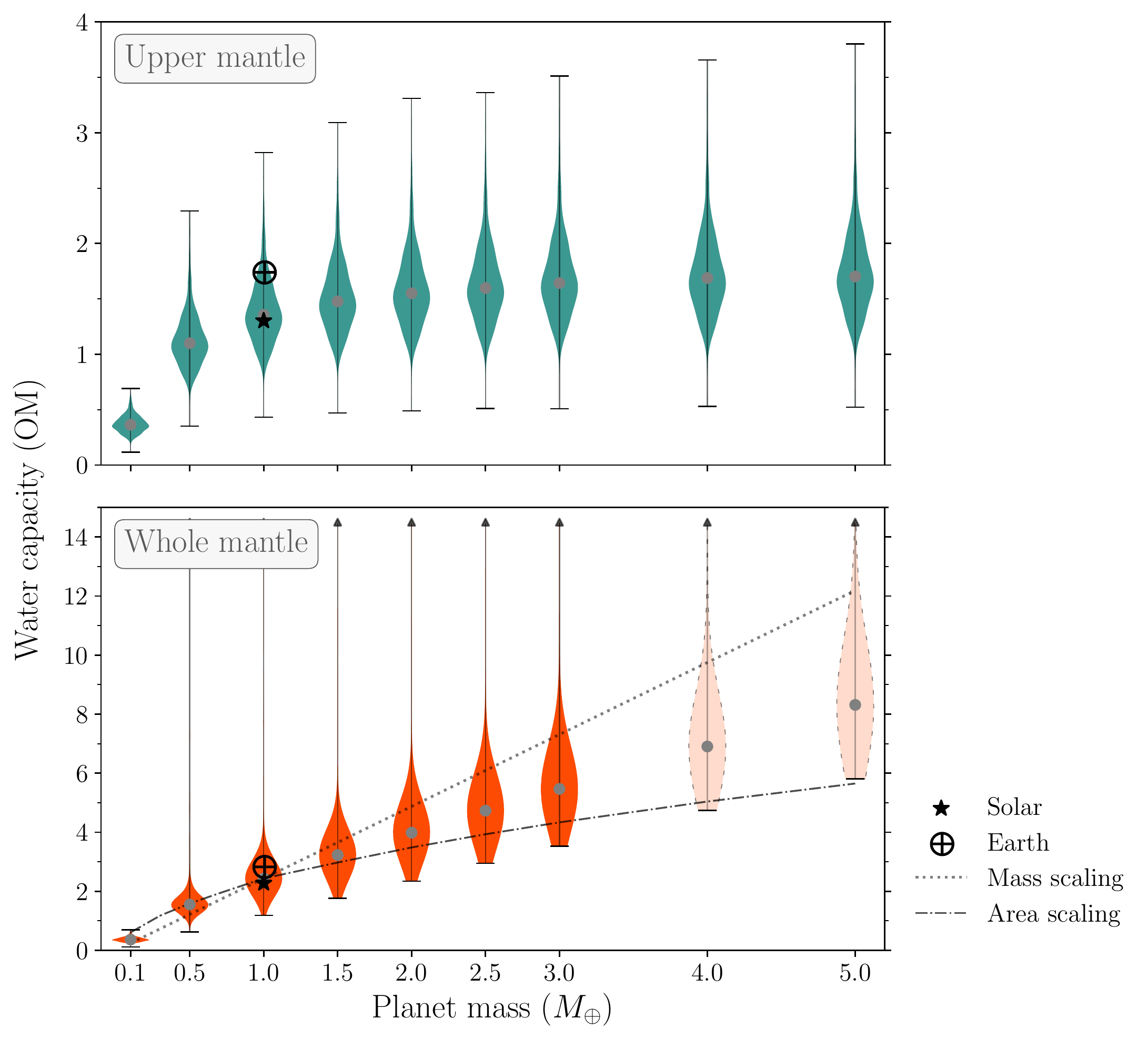}
\caption{The water storage capacities in terms of Earth ocean masses (OM; $1\,{\rm OM} = 1.335 \times 10^{21}\, {\rm kg}$), as a function of planet mass, for the upper mantle \textit{(top)} and total mantle \textit{(bottom)}. The distributions of water capacities at each planet mass arise from the variation over the entire range of host star compositions in the Hypatia Catalog ($N = 1285$). The $\oplus$ marker shows the modelled Earth value using the bulk composition of \citet{mcdonough_composition_1995}, and the star shows a planet with solar bulk composition from \citet{lodders_abundances_2009}. The dotted line scales the 1 $M_\oplus$ median by a constant water mass fraction, $\propto (M_p / M_\oplus)$, whilst the dash-dotted line scales it with planetary surface area, $\propto (M_p / M_\oplus) (g_{\rm sfc}/g_{{\rm sfc}, \oplus})^{-1}$, as suggested by \citet{cowan_water_2014}. Results are shown for a potential temperature of 1600 K and a mantle:bulk Fe of 0.113. Note that for anomalously silica-rich compositions, the total mantle water capacity can reach \textgreater1\% of the planet mass, not visible in these $y$-axis limits. Dashed violin outlines indicate planet masses which result in maximum mantle pressures above the predicted recombination of MgSiO$_3$ postperovskite with MgO or SiO$_2$ \citep{umemoto_phase_2017}, which is not included in our modelling and hence these distributions are only estimates.}
\label{fig:violin_masses}
\end{figure*}

\begin{figure*}
    \centering
    \includegraphics[width=0.9\textwidth]{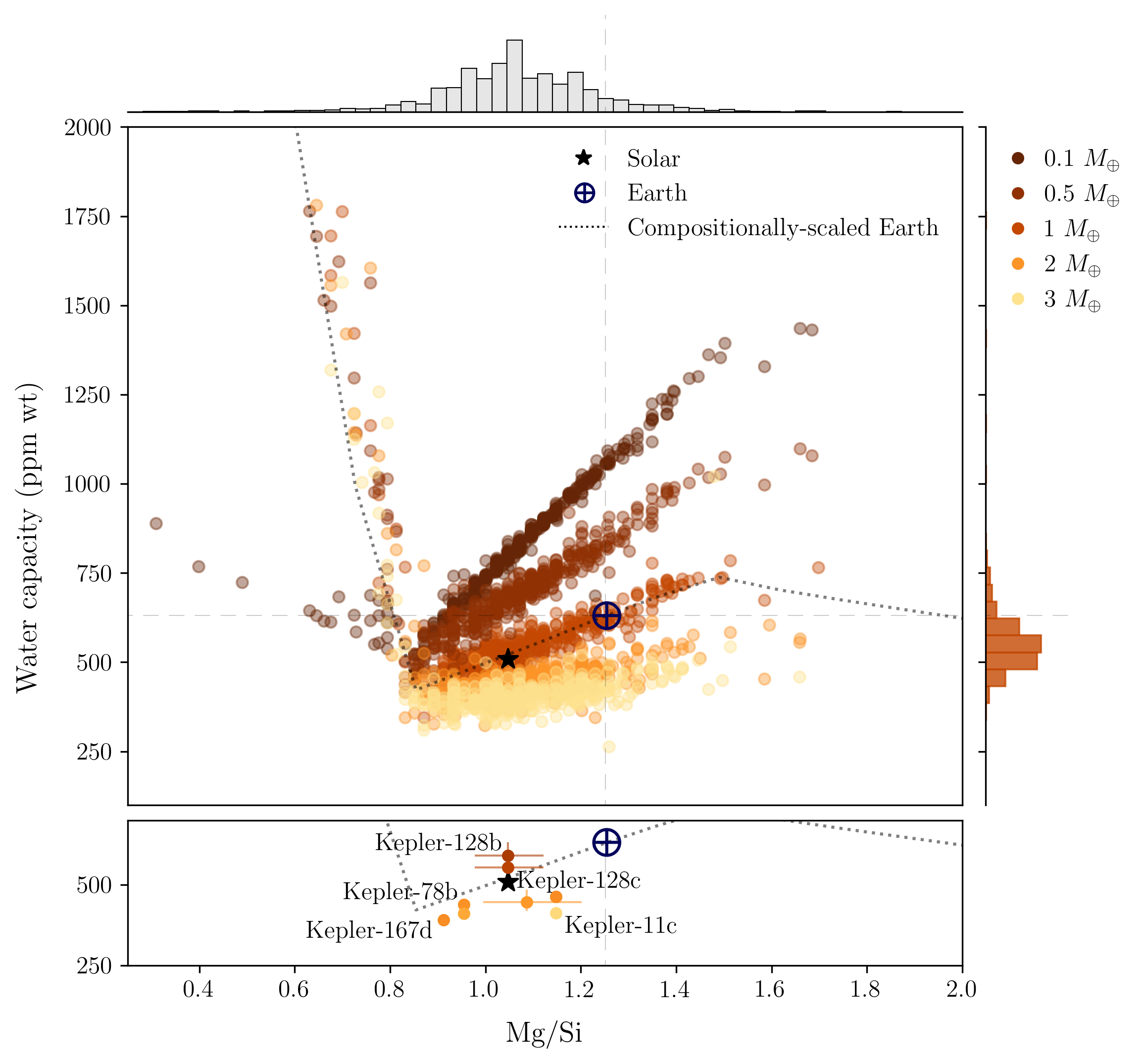}
    \caption{\textit{(Top:)} Interior water storage capacities, expressed as a mass fraction with respect to the total planet mass, of hypothetical planets having the compositions of planet-hosting stars from the Hypatia Catalog. Results are shown as a function of the Mg/Si ratio, which is the key compositional variable controlling mantle water storage capacity. Colours represent the planet mass; $2\,M_\oplus$ and $3\,M_\oplus$ planets nearly overlap. \textit{(Bottom:)} The same as above, but focused in on a selection of confirmed exoplanets using their host star compositions and median mass measurements from Hypatia (errors are propagated from the minimum and maximum stellar Mg/Si reported where multiple measurements exist). Models assume mantles with a potential temperature of 1600 K and a mantle:bulk Fe ratio of 0.113. The round symbol represents the modelled Earth value, and the dotted line represents variable Mg/Si with an otherwise Earth-like composition \citep{mcdonough_composition_1995}. The star symbol shows a planet with solar bulk composition \citep{lodders_abundances_2009}. The distributions of Mg/Si and water capacities are projected as histograms on the top and right (for $1\,M_\oplus$) axes, respectively. Note that the $y$-axis limit on water capacity crops the highest water storage capacities at the extreme low end of Mg/Si (water capacities $> 10,000$ ppm), and the $x$-axis limit on Mg/Si excludes two stars with ${\rm Mg/Si }>2$.}
    \label{fig:h2o_mgsi_scatter}
\end{figure*}

\begin{figure}
    \centering
    \includegraphics[width=0.9\columnwidth]{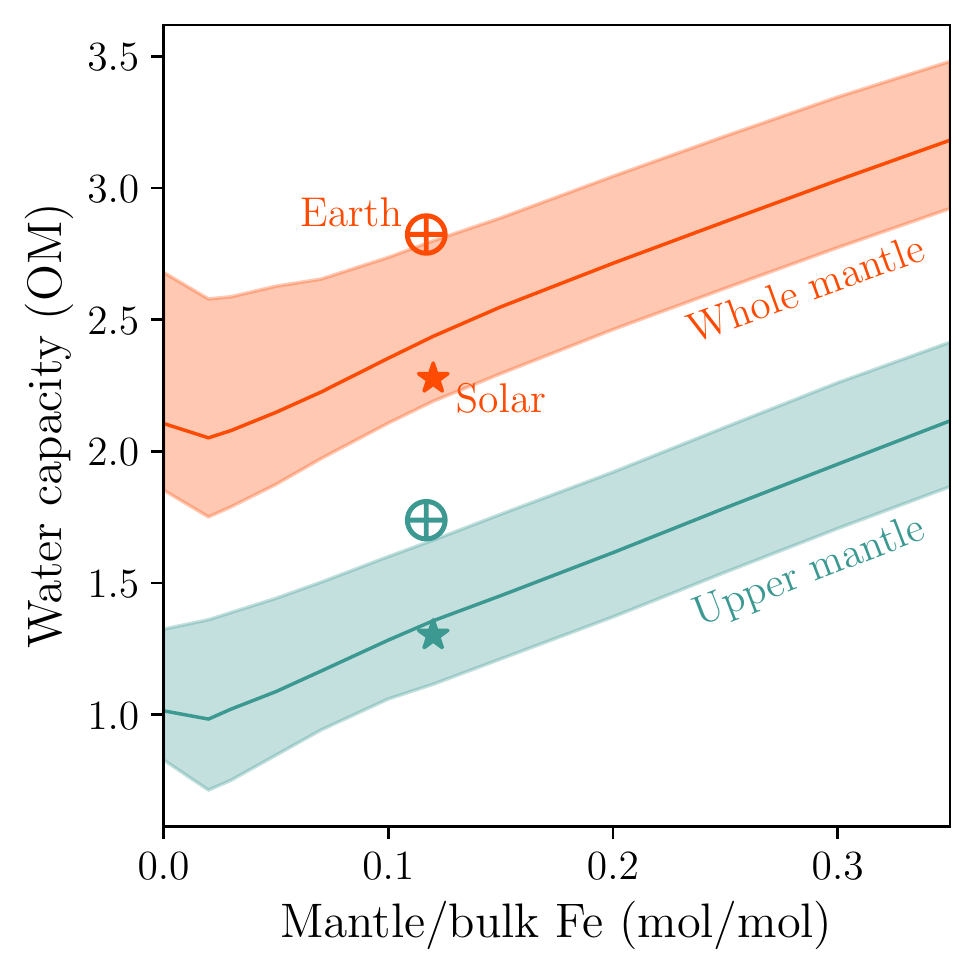}
    \caption{Total mantle (red) and upper mantle (blue) water capacities, in Earth ocean masses (OM), as a function of the molar fraction of mantle Fe to bulk Fe. Solid lines follow medians and swaths span the 1$\sigma$ distributions across compositions of planet-hosting stars from the Hypatia Catalog. Results are shown for a potential temperature of 1600 K and a planet mass of $1\,M_\oplus$. The round symbol represents the modelled Earth value using its measured composition \citep{mcdonough_composition_1995}, and the star symbol is for a planet with solar bulk composition \citep{lodders_abundances_2009}.}
    \label{fig:x_Fe_dependence}
\end{figure}

\subsection{Mantle potential temperature}\label{sec:methods_temperature}

Given that several of our key water saturation parameterisations are temperature-dependent, the choice of $T_p$ can significantly impact the overall water storage we calculate for planetary mantles. We will consider two scenarios for mantle potential temperature:

\begin{enumerate} 
\item The \textit{old, cold scenario}, set at $T_p = 1600\,{\rm K}$, nominally represents an evolved planet which has been able to cool efficiently. To distinguish the direct effect of planet mass on water capacity, we use a fixed $T_p$ for all planets, despite expecting more massive planets to have hotter interiors as a rule \citep[e.g.,][]{valencia_internal_2006}. Most of our results are shown for this cold endmember case, to emphasise a comparison with modern Earth and to isolate the effect of changing mantle mineralogy.

\item The \textit{young, hot scenario} leverages the fact that the top of a primordial mantle, having just finished crystallising from the bottom up \citep{elkins-tanton_linked_2008}, will be near its solidus temperature. It is this $T_p$ which then defines the initial $w$ condition of the saturated mantle. Further, because water saturation generally decreases with temperature, this hot limit better represents the true maximum water content of stagnant-lid planets; which, because they are not easily able to recycle water back into their interiors, will have internal water inventories at or below this initial condition. The precise solidus will be compositionally-dependent but is not currently resolved in our models \citep[see also section \ref{sec:discussion-temperature}]{hirschmann_mantle_2000}. Rather, we approximate a constant hot $T_p = 1900\,{\rm K}$. This value is around the potential temperature calculated by \citet{miyazaki_wet_2022} at which the melt-dominated layer of nominal Earth's solidifying magma ocean reaches zero thickness. 

\end{enumerate}

\section{Results}\label{sec:results}

\subsection{Compositionally-dependent mantle water capacities}

The modal mineralogy of a planet's mantle will depend strongly on its molar ratio of Mg to Si, the two most abundant refractory elements in the mantle. The largest variations in mineralogy, in response to changing Mg/Si, occur in the upper mantle. At Mg/Si around unity, (Mg,Fe)$_2$SiO$_4$ (ol polymorphs) and (Mg,Fe)$_2$Si$_2$O$_6$ (opx and hcpx) occur in roughly equal proportions; increasing Mg/Si favours (Mg, Fe)$_2$SiO$_4$ and decreasing Mg/Si favours opx, cpx and garnet. Below a critical value of Mg/Si ($\sim$0.8), however, no ol is stable at all, and the excess Si starts to appear as \ce{SiO2} polymorphs.

In the lower mantle, increasing Mg/Si favours (Mg,Fe)O (fp) at the expense of (Mg,Fe)SiO$_3$ (pv), and decreasing Mg/Si favours the converse. Again, \ce{SiO2} appears in appreciable quantities below a critical value of Mg/Si.

Fig. \ref{fig:mgsi_modality} illustrates the mineral mode changes and resulting water capacity profiles from simply varying Mg/Si across the typical range in our stellar sample, 0.72--1.41 (95\% of stars). At the high end of this range, the dominating wad and ring phases lead to a large capacity for water storage in the MTZ. With decreasing Mg/Si, the proportions of ol, wad, and ring diminish, and the water capacity of the MTZ correspondingly decreases as well. These overall shapes of the water saturation profile around the MTZ are robust to the details of water capacity parameterisations, given that $D^{\rm wad}_{\rm ol} \gg D^{\rm ol}_{\rm opx}$ \citep{bercovici_whole-mantle_2003, dong_constraining_2021, andrault_mantle_2022}. 


Although Fig. \ref{fig:mgsi_modality} does not show most of the lower mantle, decreasing Mg/Si also shrinks the proportion of ppv relative to pv by deepening the pv-ppv transition pressure. Given that $c_{\rm ppv} > c_{\rm pv}$ \citep[in the presence of Al;][]{townsend_water_2016}, this lower mantle effect also contributes to the decrease in overall water capacity with decreasing Mg/Si.

Upon a further decrease of Mg/Si beyond $\sim$0.8, the trend of decreasing water capacity with decreasing Mg/S starts to break. Here, \ce{SiO2} polymorphs can now form sizeable proportions of the mantle volume, in a quite un-Earth-like regime (note that allowing Si to partition into the core would diminish the occurrence rate of planets with low Mg/Si). Our model adopts $c_{\rm stv}$ comparable to wad and ring \citep{panero_hydrogen_2004, litasov_high_2007}. If stv indeed has such potential as a water reservoir, then the lower mantle's sheer volume means that even small proportions of stv (<10 wt\%) could contribute many Earth oceans of water storage capacity. This effect is especially distinctive for larger planets, as we discuss in the next section.


\subsection{Scaling mantle water capacity with planet mass}

Having demonstrated how the MTZ features prominently in mantle water capacity profiles across a range of compositions, we explore the consequences of this fact on $w$-$M_p$ scaling relationships. High-water-capacity phases characteristic to the MTZ are only stable in a thin layer between about 14 and $24\,{\rm GPa}$---pressure limits which hold, with only small variation, across most planet compositions. 
The further, less obvious fact is that as $M_p$ increases beyond $1\,M_\oplus$, the ring-pv transition pressure does not underlay markedly more mass in the overlying upper mantle, even with the increased planet surface area (Fig. \ref{fig:mass_um}). In other words, beyond $1\,M_\oplus$, upper mantle masses become nearly independent of $M_p$ ($\propto M_p^{0.15}$ if $M_p > 1 \, M_\oplus$) 

Note that Mars-sized ($\sim$0.1$\,M_\oplus$) planets will not be deep enough to stabilise pv. Such planets would not have bona fide lower mantles; only upper mantle phases could contribute water capacity.


Fig. \ref{fig:violin_masses} presents the water capacity distributions we predict over chemically-diverse exoplanets as a function of their mass, shown here for an ``old and cold'' $T_p = 1600\,{\rm K}$ and an Earth-like \coreeff. As the upper mantle mass asymptotes with increasing $M_p$, so does the upper mantle water mass at water saturation---approaching a median of about 1.7 Earth ocean masses (OM; $1\,{\rm OM} = 1.335\times 10^{21}\,{\rm kg}$) at this $T_p$ (Fig. \ref{fig:violin_masses}; top panel). Mineralogical variability leads to an interquartile range of $\sim$0.4~OM for $M_p > 1\, M_\oplus$. 

Meanwhile, the water capacity of the whole mantle (Fig. \ref{fig:violin_masses}, bottom panel) necessarily increases with $M_p$. Despite the uncertainty due to poorly-constrained water saturations of lower mantle phases (see section \ref{sec:results_sensitivity}), the slope of $w_m$ with $M_p$ is credible. It falls, as expected, between a constant scaling of Earth's water capacity with $M_p$ and with planet surface area: 

\begin{enumerate}
\item The median $w_{\rm m}$ should be lower, and have a shallower slope with $M_p$, than the value implied by the scaling that treats the whole mantle as having a constant water saturation: $(w_{\rm m}/w_{\rm m, \oplus}) \propto (M_p / M_\oplus)$ (Fig. \ref{fig:violin_masses}, dotted line). This is because we know that water saturation peaks sharply at the MTZ, which represents less and less of the overall mantle mass at larger $M_p$.


\item As suggested by \citet{cowan_water_2014}, water capacity would scale in the same way as the surface area if it were focused \textit{only} into a thin shell in the upper mantle; that is, $(w_{\rm m} / w_{\rm m, \oplus}) \propto (M_p / M_\oplus) (g_{\rm sfc}/g_{\rm sfc, \oplus})^{-1}$ (Fig. \ref{fig:violin_masses}, dashed line). The true median $w_m$ should be higher than the value implied by this scaling because the rest of the mantle has nonzero water capacity. 

\end{enumerate}

Whilst the scaling in \textit{(i)} is not appropriate around 1 $M_\oplus$, $w_{\rm m}$ does indeed increase roughly linearly with $M_p$ at higher $M_p$, where the upper mantle is a less significant fraction of the whole mantle (and given our model where $c_{\rm ppv}$ is more or less constant with $p$). Between 1 and 1.5 $M_\oplus$, the integrated lower mantle water capacity first begins to outstrip $w_{\rm um}$; by 3 $M_\oplus$ the lower mantle can hold about 2.5 times the water mass of the upper mantle. 


Finally, also intriguing in Fig. \ref{fig:violin_masses} is the suggestion that an Earth-like Mg/Si results in a higher interior water capacity with respect to the average Mg/Si measured in FGKM stars. Earth's mantle Mg/Si places it at the 85th percentile of $w_{\rm m}$ for planets with the same mass and Fe partitioning---this is due to larger fractions of wad and ring in Earth's MTZ than is typically predicted for planets. However, as discussed in section \ref{sec:discussion-sio2}, the Mg/Si abundances of stars may be lower than the composition of their associated rocky planet mantles if some fraction of a planet's inherited bulk Si were commonly locked in the metal core during formation. Indeed, the raw solar Mg/Si is actually slightly below the stellar average in Hypatia; a planet with Si-Mg-Fe-Ca-Al composition at exactly the solar value and an Earth-like core fraction would have $w_{\rm m}$ at the 31st percentile.

Evaluating our parameterisation for an Earth-like composition gives an MTZ water capacity of 1.25 OM and an olivine-bearing mantle water capacity of 0.12 OM, if we use the same potential temperature ($T_p = 1650\,{\rm K}$) as \citet{andrault_mantle_2022}---corresponding to a wetter MTZ and drier olivine-bearing mantle compared to that work, which finds 0.6 OM and 0.31 OM, respectively, and a drier upper mantle overall.

\subsection{Interior water mass fractions of rocky planets}

Fig. \ref{fig:h2o_mgsi_scatter} shows these same interior water capacities not as an absolute water mass, but as a mass fraction of water with respect to $M_p$. Although more massive planets can clearly sequester more water, these water masses corresponds to much lower fractions of the overall planet mass. A planet with Earth's bulk composition has a mantle water capacity of $\sim$1000~ppm at 0.1 $M_\oplus$, but the mantle of a $3\,M_\oplus$ planet of the same composition can only contain $\sim$400~ppm. The corollary of this is that if these two planets accreted the same mass fractional water budget, the larger planet would be able to store less of this water in its mantle. This point carries notable implications for the frequency of ocean-covered planets and is discussed in section \ref{sec:discussion-waterworld}.  

There is a water capacity trend with bulk composition at all planet masses. In the typical range of stellar Mg/Si, water capacity increases with increasing Mg/Si because more wad and ring are stabilised at the expense of gt in the MTZ region (Fig. \ref{fig:mgsi_modality}). The slope of water capacity with Mg/Si flattens at higher $M_p$ because the contribution of the $\sim$14--24-GPa MTZ becomes less important than the lower mantle. That is, the water capacities of pv, ppv, and fp are similar, so their changing proportions due to changes in bulk composition have little effect on lower mantle water storage capacity.

To highlight the compositional dependence of mantle water storage capacity, we also isolate the effect of changing Mg/Si on its own for an otherwise Earth-like bulk composition (Fig. \ref{fig:h2o_mgsi_scatter}, black dotted line). Again, there is an evident increase in water capacity around the Mg/Si range near Earth's bulk composition. However, we see two kinks around the extrema of Mg/Si $\lesssim 0.8$ and Mg/Si $\gtrsim 1.5$, where key changes in the modal mineralogy occur. In the former shift---the extreme low end---all Mg has been used in forming pyroxenes, and excess Si is left as the silica polymorphs qtz, coes, stv, and seif. Silica-rich mantles can attain very high water contents, due to the high water capacity of stv we adopt here (section \ref{sec:methods_sat}). Mars-sized (0.1 $M_\oplus$) planets do not have deep-enough mantles to host significant masses of stv, so the associated water capacity increase at low Mg/Si is only moderate. Meanwhile, for Mg/Si $\gtrsim1.5$, there is excess Mg beyond that which is incorporated into wad and ring, so dry fp appears even in the MTZ.

For illustration, Fig. \ref{fig:h2o_mgsi_scatter} also shows our calculated interior water mass fractions for a selection of known exoplanets, given their mass and host star composition measurements. 

\subsubsection{Sensitivity to water saturation parameterisation}\label{sec:results_sensitivity}
A large part of the uncertainty on the interior water capacities of planets near $1\,M_\oplus$ comes from the unknown water storage of pv. \citet{dong_constraining_2021} tested the sensitivity of an Earth-mass, Earth-composition mantle water capacity to  $D^{\rm ring}_{\rm pv} \in [7, 23]$ \citep{inoue_water_2010}, and simultaneously to the regression uncertainty on their fit parameters, finding 25th- and 75th- percentile water capacities at around $\sim$10--20\% away from the 50th percentile value. Predictions of mantle water capacities for any bigger planet would be more uncertain, predominantly because the water capacity of postperovskite, and its pressure dependence, are so poorly constrained. Postperovskite could reasonably hold more water with increasing pressure---as does olivine and stishovite \citep{panero_hydrogen_2004, dong_constraining_2021}---in which case massive planets would be effective at sequestering deep water. Nevertheless, these deep mantle water reservoirs might not be communicating efficiently with the planetary surface, especially if a low $c_{\rm pv}$ at the top of the lower mantle creates a bottleneck to water circulation (see section \ref{sec:discussion-dynamics}).

\subsection{Mineralogical and structural effects of planetary Fe partitioning}\label{sec:results_fe}

So far, we have shown results for planets with constant \coreeff. Changing how Fe is partitioned between the mantle and the core of a planet has several distinct effects on the interior water capacity (Supplementary Figs. S1--S3):

\begin{enumerate}
    \item A more massive core (with a resultant lower CMB pressure) will shift the entire mass profile of the mantle upwards, and steepen the pressure gradient, with the consequence of less mass underlying the ring-pv transition, and a lower UM mass \citep{unterborn_scaling_2016, unterborn_pressure_2019}.
    \item A more FeO-rich mantle will also have a steeper pressure gradient.
    \item A more FeO-rich mantle will have a higher water concentration at saturation, all else equal, because this composition favours olivine polymorphs over gt in the UM, produces a shallower ol-wad transition in the UM, and produces a shallower pv-ppv transition in the LM.
\end{enumerate}

If we move iron from the core to the mantle (as FeO), the net effect is a slightly higher UM mass, and a higher average water saturation in ppm wt for both the UM and LM. These effects amplify each other to result in a larger water capacity at larger \coreeff~(Figure \ref{fig:x_Fe_dependence}).

The other way to change CMF and mantle FeO content in our model, which conserves $M_p$ and bulk composition, is to have a more Fe-rich star: extra iron everywhere raises both the CMF and the mantle FeO content. In contrast to changing \coreeff, these two effects have competing consequences for water capacities---a lower UM mass but a higher water saturation---with the net effect being a slight decrease (more so at higher $M_p$). 

A corollary of the results described above is that mantle water capacities depend more strongly on the Mg/Si ratio than on the Fe/Si ratio of the system. This analysis does not account for possible direct effects of Fe on mineral water saturation, which, except for olivine, are not easily resolvable with the current experimental data \citep{dong_water_2022}.

\subsection{Results summary}

Table \ref{tab:all_results} lists our results for the upper and total mantle water capacity distributions as a function of $M_p$, $T_p$, and \coreeff. Although the results presented so far have focused on an old and cold scenario of $T_p = 1600\,{\rm K}$, we have suggested that the hot temperature profile, realised immediately after magma ocean crystallisation, is the relevant temperature that locks in the water content of the interior---at which the solid seals itself from the overlying water (see section \ref{sec:discussion-temperature}). Hence Table \ref{tab:all_results} also reports results at $T_p = 1900\,{\rm K}$.


\begin{table*}
\centering
\caption{Mantle water capacity medians across stellar compositions from the Hypatia Catalog, and 2$\sigma$ widths in parentheses, as a function of planet mass ($M_p$), potential temperature ($T_p$), and molar mantle:bulk iron ratio (\coreeff). $1\,{\rm OM} = 1.335 \times 10^{21}\,{\rm kg}$. \label{tab:all_results}}
\footnotesize

\begin{tabular}{@{} c c l l l l l @{}}
\toprule
& & \multicolumn{5}{c}{\coreeff} \\ \cmidrule(lr){3-7}
$T_p\,{\rm (K)}$ & $M_p\,(M_\oplus)$ & 0.30 & 0.20 & 0.12 & 0.05 & 0.00 \\
\midrule
\noalign{\vskip 1mm}

\multicolumn{7}{c}{\textbf{Whole mantle water capacity (OM)}} \\
\multirow{10}{*}{1600} & 0.1 & 0.60 (0.40) & 0.47 (0.35) & 0.37 (0.32) & 0.29 (0.30) & 0.24 (0.29) \\
 & 0.3 & 1.53 (0.89) & 1.30 (0.94) & 1.11 (0.96) & 0.92 (1.04) & 0.82 (1.16) \\
 & 0.5 & 2.05 (1.14) & 1.79 (1.37) & 1.56 (2.01) & 1.33 (2.77) & 1.29 (3.14) \\
 & 1.0 & 3.03 (1.61) & 2.71 (3.01) & 2.44 (5.16) & 2.15 (7.14) & 2.11 (8.05) \\
 & 1.5 & 3.87 (1.97) & 3.53 (4.41) & 3.23 (7.35) & 2.92 (10.24) & 2.86 (11.49) \\
 & 2.0 & 4.66 (2.27) & 4.30 (5.83) & 3.99 (9.44) & 3.66 (12.98) & 3.61 (14.48) \\
 & 2.5 & 5.40 (2.61) & 5.04 (7.23) & 4.73 (11.47) & 4.41 (15.61) & 4.34 (17.29) \\
 & 3.0 & 6.13 (3.00) & 5.78 (8.63) & 5.47 (13.43) & 5.13 (18.09) & 5.06 (19.86) \\
 & 4.0 & 7.56 (3.80) & 7.22 (11.36) & 6.90 (17.25) & 6.57 (23.02) & 6.50 (25.22) \\
 & 5.0 & 8.96 (4.95) & 8.62 (13.90) & 8.31 (21.04) & 7.99 (27.81) & 7.91 (29.88) \\
\hline
\multirow{10}{*}{1900} & 0.1 & 0.17 (0.11) & 0.13 (0.10) & 0.11 (0.11) & 0.09 (0.12) & 0.09 (0.14) \\
 & 0.3 & 0.54 (0.31) & 0.48 (0.85) & 0.43 (1.35) & 0.39 (1.75) & 0.39 (2.31) \\
 & 0.5 & 0.83 (0.54) & 0.76 (1.89) & 0.70 (3.12) & 0.66 (4.32) & 0.65 (5.57) \\
 & 1.0 & 1.50 (1.07) & 1.41 (4.13) & 1.35 (7.05) & 1.30 (9.67) & 1.29 (11.18) \\
 & 1.5 & 2.16 (1.61) & 2.08 (5.69) & 2.01 (9.44) & 1.96 (13.25) & 1.95 (15.02) \\
 & 2.0 & 2.82 (2.11) & 2.74 (7.12) & 2.67 (11.68) & 2.63 (16.30) & 2.62 (18.40) \\
 & 2.5 & 3.48 (2.60) & 3.40 (8.51) & 3.34 (13.79) & 3.29 (19.06) & 3.28 (21.71) \\
 & 3.0 & 4.14 (3.10) & 4.07 (9.88) & 4.01 (15.87) & 3.96 (21.63) & 3.95 (24.56) \\
 & 4.0 & 5.47 (4.10) & 5.40 (12.55) & 5.34 (19.75) & 5.30 (26.66) & 5.30 (29.59) \\
 & 5.0 & 6.79 (5.08) & 6.40 (35.00) & 6.68 (23.53) & 6.64 (31.35) & 6.65 (34.84) \\
\hline
\multicolumn{7}{c}{\textbf{Upper mantle water capacity (OM)}} \\
\multirow{10}{*}{1600} & 0.1 & 0.58 (0.41) & 0.46 (0.35) & 0.36 (0.32) & 0.28 (0.30) & 0.24 (0.29) \\
 & 0.3 & 1.30 (0.91) & 1.07 (0.87) & 0.89 (0.82) & 0.71 (0.81) & 0.61 (0.84) \\
 & 0.5 & 1.60 (1.11) & 1.32 (1.07) & 1.10 (1.01) & 0.88 (1.02) & 0.82 (0.95) \\
 & 1.0 & 1.95 (1.37) & 1.62 (1.33) & 1.36 (1.26) & 1.09 (1.27) & 1.01 (1.19) \\
 & 1.5 & 2.12 (1.49) & 1.76 (1.44) & 1.48 (1.36) & 1.19 (1.39) & 1.11 (1.28) \\
 & 2.0 & 2.22 (1.57) & 1.85 (1.52) & 1.55 (1.45) & 1.25 (1.45) & 1.17 (1.38) \\
 & 2.5 & 2.28 (1.61) & 1.90 (1.56) & 1.60 (1.48) & 1.29 (1.50) & 1.22 (1.41) \\
 & 3.0 & 2.35 (1.65) & 1.95 (1.61) & 1.64 (1.51) & 1.33 (1.56) & 1.25 (1.45) \\
 & 4.0 & 2.40 (1.70) & 1.99 (1.65) & 1.69 (1.56) & 1.36 (1.58) & 1.28 (1.49) \\
 & 5.0 & 2.41 (1.70) & 2.01 (1.66) & 1.70 (1.60) & 1.38 (1.61) & 1.29 (1.50) \\
\hline
\multirow{10}{*}{1900} & 0.1 & 0.16 (0.11) & 0.13 (0.10) & 0.11 (0.11) & 0.09 (0.12) & 0.09 (0.14) \\
 & 0.3 & 0.37 (0.27) & 0.31 (0.27) & 0.26 (0.27) & 0.23 (0.32) & 0.22 (0.37) \\
 & 0.5 & 0.46 (0.33) & 0.38 (0.34) & 0.33 (0.34) & 0.29 (0.40) & 0.27 (0.45) \\
 & 1.0 & 0.56 (0.41) & 0.47 (0.42) & 0.40 (0.42) & 0.36 (0.49) & 0.33 (0.54) \\
 & 1.5 & 0.61 (0.45) & 0.51 (0.46) & 0.44 (0.46) & 0.39 (0.53) & 0.36 (0.57) \\
 & 2.0 & 0.64 (0.47) & 0.53 (0.48) & 0.46 (0.48) & 0.41 (0.57) & 0.38 (0.63) \\
 & 2.5 & 0.66 (0.48) & 0.55 (0.49) & 0.48 (0.49) & 0.42 (0.59) & 0.39 (0.62) \\
 & 3.0 & 0.67 (0.51) & 0.57 (0.51) & 0.49 (0.52) & 0.43 (0.58) & 0.40 (0.65) \\
 & 4.0 & 0.68 (0.51) & 0.58 (0.52) & 0.50 (0.53) & 0.44 (0.61) & 0.40 (0.65) \\
 & 5.0 & 0.69 (0.51) & 0.65 (0.63) & 0.51 (0.54) & 0.45 (0.61) & 0.41 (0.70) \\
\hline


\end{tabular}
\end{table*}

\section{Discussion}\label{sec:discussion}

\subsection{Implications and consequences}

\subsubsection{The occurrence rate of ocean planets}\label{sec:discussion-waterworld}

A key question motivating this work is how water partitions between a planet's interior and surface. This partitioning determines whether planets become fully-inundated ocean planets, or can maintain some fraction of exposed land. We have focused on one specific process out of many that feed into a planet's interior-surface water partitioning: the ability of mantle minerals to store any water with which the planet is endowed during its formation. We have shown that higher-mass rocky planets can sequester proportionally less water in their interiors, down to $\sim$300--400 ppm of the planet mass at $M_p \gtrsim 3\,M_\oplus$.

A detailed understanding of ocean planet propensity must unite all processes that govern interior-surface water partitioning. Ultimately, whether or not planets have dry land depends on the interplay of \textit{(i)} total water budgets, \textit{(ii)} water redistribution between interior and surface reservoirs, and \textit{(iii)} the theoretical capacities of these surface (i.e., ocean basin volume) and interior reservoirs. 

Estimating the initial water budget of a planet \textit{(i)} requires knowing the amount of water in planetary building blocks and the ability of the growing planet to ingas nebular hydrogen. These variables depend in turn on the planetary formation time- and space-scales, as well as on $M_p$ \citep[and references therein]{sharp_nebular_2017, unterborn_inward_2018, kimura_formation_2020, ohtani_hydration_2021}. Planetary accretion simulations, although stochastic, produce rocky planets in the classical habitable zone commonly forming with $\gtrsim$10 OM water \citep{raymond_making_2004, raymond_high-resolution_2006, zain_planetary_2018}. The oxidation of ingassed nebular H could alone contribute $\gtrsim$1000~ppm water, without needing to capture icy planetesimals, depending on the stellar mass and nebular lifetime \citep{olson_nebular_2019, kimura_predicted_2022}. The water contents of enstatite, ordinary, and carbonaceous chondrites (\textgreater 1000--10,000~ppm), materials thought to sample planetary building blocks, easily exceed the mantle water capacities of Earth-mass planets and above \citep{abe_water_2000}. Note that late delivery of water by comets is not expected to contribute a significant amount of water, at least in the solar system \citep{morbidelli_source_2000}. Overall, it appears quite possible for planets to accrete water beyond that which their mantles can hold.

Estimating water redistribution \textit{(ii)} is more tractable in the stagnant lid regime. Here, with no mantle return flux, redistribution is controlled by volcanic outgassing and is tied to the thermal history of the planet \citep[e.g.,][]{ortenzi_mantle_2020}. The persistence of early oceans extruded from a cooled magma ocean is also strongly coupled with the planetary climate \citep[e.g.,][]{elkins-tanton_formation_2011, miyazaki_wet_2022}. 

Constraints on the interior reservoir capacity \textit{(iii)} are covered in this work. The potential size of the \textit{surface} reservoir depends on topography, as was shown in \citet{guimond_blue_2022}, who found that more massive planets are virtually topographically featureless and able to be flooded by a relatively smaller ocean volume.

Based our prediction of maximum mantle water mass fractions decreasing roughly as $M_p^{-0.23}$, and our previous work finding the maximum surface water mass fraction (based on the bare-minimum topography) to decrease as $M_p^{-0.75}$ \citep{guimond_blue_2022}, we speculate that land and oceans are less likely to coexist on massive rocky planets. At the end of the magma ocean stage, any water in excess of their limited solid mantle water capacities would immediately contribute to deep oceans (or thick steam atmospheres)---or, if this non-sequestered water escapes under high stellar irradiation, the planets are irreversibly desiccated. This speculation is tied to the low water capacity of pv and ppv adopted here, and would need to be re-evaluated if, for example, future experiments suggest that these deep mantle silicates can store significantly more water at higher pressures.


\subsubsection{Consequences and corollaries of a temperature dependence of mantle water storage capacity}
\label{sec:discussion-temperature}

\begin{figure}
    \centering
    \includegraphics[width=\columnwidth]{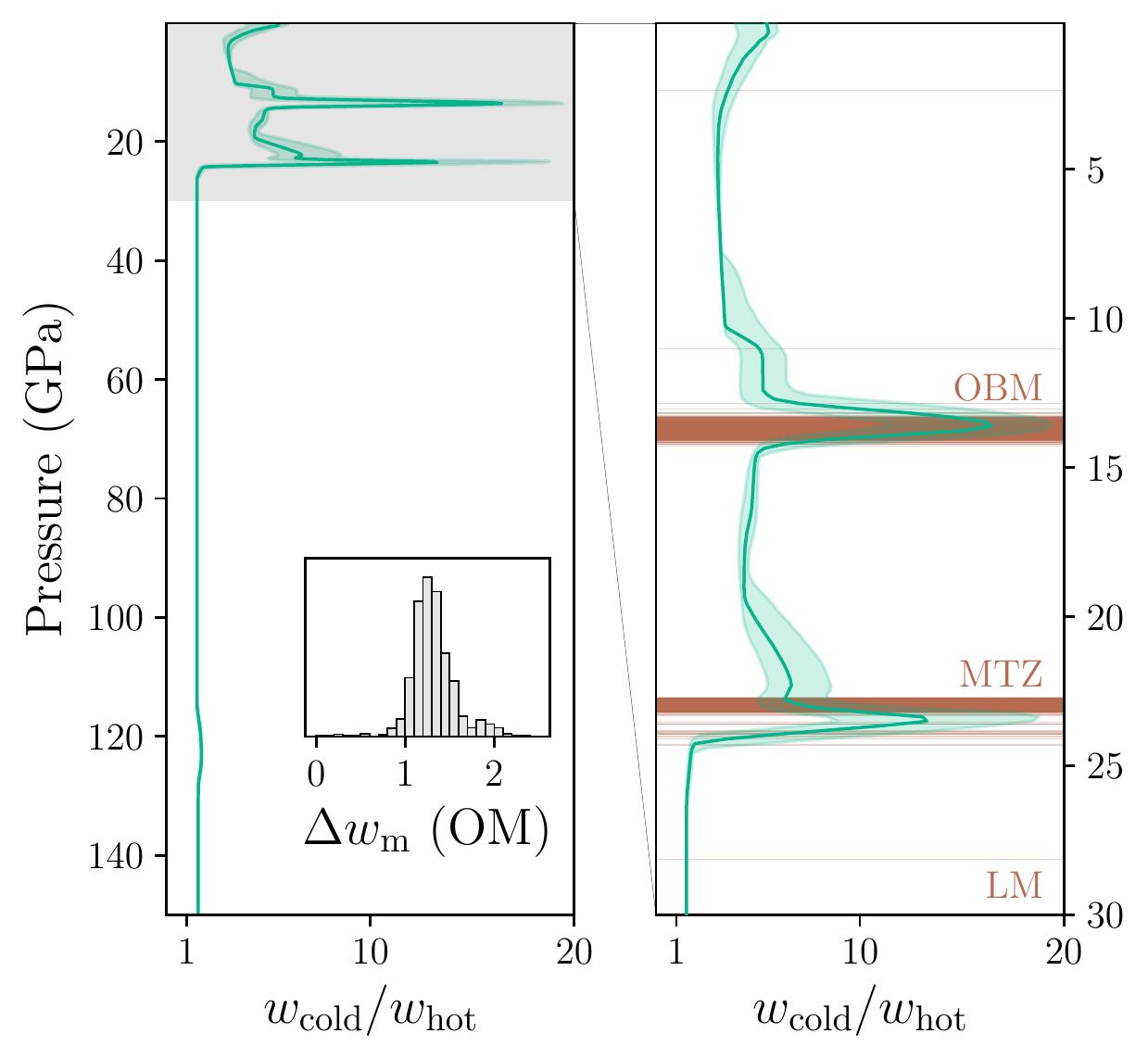}
    \caption{The temperature dependence of mantle water storage capacity. Mass ratios of mantle water capacities ($w_{\rm cold}/w_{\rm hot}$) for mantles with potential temperatures of $1600\,{\rm K}$ and $1900\,{\rm K}$ are shown in two profiles, for the whole mantle \textit{(left)} and in detail to the top of the lower mantle \textit{(right)}. The swath spans the 1$\sigma$ distribution across compositions of planet-hosting stars from the Hypatia Catalog; the solid blue line shows the median. As the mantle cools, the water capacity increases for all compositions shown. Horizontal lines show the first appearances of wadsleyite and of perovskite with increasing pressure, demarcating the respective mantle regions: olivine-bearing mantle (OBM), mantle transition zone (MTZ), and lower mantle (LM). The largest values of $w_{\rm cold}/w_{\rm hot}$ are around these phase transitions, which shift $<1\,{\rm GPa}$ between the two $T_p$. This profile pertains to a $1\,M_\oplus$ planet. \textit{(Inset:)} Distribution of the difference in the whole mantle water storage capacity, in Earth ocean masses (OM), between mantle potential temperatures of $1600\,{\rm K}$ and $1900\,{\rm K}$. Not shown in the histogram axis limits are the results for anomalously silica-rich planets, the water storage capacities of which may increase with temperature.}
    \label{fig:sat_diff_T}
\end{figure}

As the mantle cools, we expect NAMs to hold more water at saturation \citep{keppler_thermodynamics_2006}---only stishovite behaves conversely in our parameterisation. For a closed system (e.g., the mantle of a stagnant lid planet undergoing no water recycling to its interior), however, the mantle water content cannot increase as the system cools, even if the minerals could in principle hold more water. We nonetheless explore the effect of temperature on the \textit{theoretical} maximum water storage capacity of planets, as this will be relevant for \textit{(i)} planets with an active surface-mantle water cycle (e.g., plate tectonic planets), and \textit{(ii)} showing how the temperature at the end of magma ocean solidification may set the mantle's initial water capacity.

We find that as a $1\,M_\oplus$ planet secularly cools by $300\,{\rm K}$ over time \citep[an amount by which Earth's own mantle may have cooled over its history;][]{herzberg_thermal_2010}, its capacity to store water will increase by $\sim$1--2 OM (Fig. \ref{fig:sat_diff_T}), as highlighted elsewhere \citep{shah_internal_2021, dong_constraining_2021, andrault_mantle_2022}. At the present time, we cannot rectify any temperature dependence of water saturation in pv or ppv with the available data. Theoretical calculations by \citet{hernandez_incorporation_2013} find $D^{\rm ring}_{\rm pv}$ decreasing with temperature, which would mute any $T$-dependence of $c_{\rm pv}$ given that $c_{\rm ring}$ is also lower at high $T$.

The direct thermodynamic effect of cooling on water capacity is mildly stronger than the effect of variations in upper mantle mineralogy: in Fig. \ref{fig:sat_diff_T} the median ratio $w_{\rm cold}/w_{\rm hot}$ surpasses the width of the distribution of mantle water storage capacity from predicted mantle mineralogies (cf. Fig. \ref{fig:violin_masses}). Note, however, that the largest ratios ($w_{\rm cold}/w_{\rm hot}>10$) are seen around the ol-wad transition and the ring-pv transitions. Here, higher $T_p$ shifts the pressures of these transitions deeper by $\lesssim1\,{\rm GPa}$. Much higher water capacities in wad and ring effectively create a large $w_{\rm cold}/w_{\rm hot}$ ratio around these pressures.
\medskip

Although the fact that temperature exerts a first-order control on mineral water saturation should elicit caution, we can leverage certain deterministic consequences of the thermal evolution of planets to apply our results further.



\medskip

\textbf{Patterns in $T_p$ across planets}---the mantle temperatures of a planet at thermal quasi-steady state relate predictably to its mass and tectonic mode. Whilst a $T_p$ of $\sim$1600$\,{\rm K}$ is appropriate for an Earth-mass planet in a plate tectonic regime, more massive planets will run hotter.
The volume of a sphere increases faster than its surface area, so the total volumetric heating (e.g., through radioactive decay) increases faster than does the heat lost from the top of the mantle. 
Hotter mantle temperatures at larger $M_p$ means lower water solubilities for the constituent minerals (Fig. \ref{fig:sat_diff_T}). The result is an even sharper decline in water storage capacity with increasing $M_p$ compared to Fig. \ref{fig:h2o_mgsi_scatter}.

Similarly, we would predict that water capacities should be generally lower for all planets in the stagnant lid regime. In this geodynamic mode, planets exhibit less-efficient surface heat loss and accordingly run hotter than their similarly-aged plate tectonic counterparts \citep[e.g.,][]{kite_geodynamics_2009}. 

Thermal history models are not free of their own uncertainty, however. Predicted mantle temperatures in mature planets can vary on the order of $100\,{\rm K}$ depending on the activation energy of viscosity, which itself is set by mineralogy. The present work does not implement the extra complexity of an informed $T_p$, in particular because compositional effects on viscosity are potentially significant and not easy to constrain. As \citet{spaargaren_influence_2020} discuss, we would generally expect less-viscous lower mantles at higher Mg/Si because fp is weaker than pv, and the overall viscosity is controlled by the weaker phase \citep{girard_shear_2016}; higher opx modes in the upper mantle may also cause rheological weakening \citep{tasaka_rheological_2020}. Viscosities further depend on the instantaneous water content itself, whereby a water-saturated mantle should be rheologically weaker than a dry mantle \citep{karato_rheology_1993}. This discussion points to feedbacks between mantle temperature, composition, geodynamics and water storage capacity that in future studies will be important to consider.

\medskip

\textbf{Lower mantle water capacities in the past}---the fact that mantle minerals store less water with increasing $T_p$ allows us to estimate how much change can occur in a planet's interior-surface water partitioning as a function of its age. Planets necessarily cool down from their primordially-hot state with secular evolution over billions of years. Therefore, mantle water capacities will increase with age for most planets, with the result that their minimum water storage capacities are met early in their histories.

For planets with functioning return fluxes of water to the mantle (e.g., subducting plates), the net difference in total water capacity with secular mantle cooling, $\Delta w_{\rm m} (\Delta T_p)$, approximates the mass of surface water that the planet could ingest (Fig. \ref{fig:sat_diff_T}). This idea was previously explored by \citet{dong_constraining_2021, dong_water_2022} and \citet{andrault_mantle_2022}, who, as in this study, found an increase in total water capacity of 1--2 OM with mantle cooling from 1900 to 1600 K. 

For planets in the stagnant lid regime, within which water flows unidirectionally from mantle to surface, a corollary is that the initial condition and upper limit of mantle water content are one and the same, and are set by the shallow solidus temperature (section \ref{sec:methods_temperature}). 

This important role of the solidus temperature calls for better experimental data on mantle solidii of arbitrary compositions. \citet{brugman_experimental_2021} showed a significant depression of the solidus temperature when Ca/Al is increased beyond that typical for peridotites; however, this disparity seems to disappear at $1\,{\rm GPa}$. Meanwhile, the solidus temperature can change by $\gtrsim$100~K due to varying mantle alkali (Na and K) abundances \citep{hirschmann_mantle_2000}. The issue is that these elements are moderately volatile and are therefore likely to fractionate with respect to their stellar relative abundances, via processes in the protoplanetary disk and the growing planet itself. Consequently, it is difficult to be deterministic about Na or K abundances in exoplanets from looking at stellar relative abundances alone \citep[see][]{wang_detailed_2022}. Hence exoplanet mantle solidi may be difficult to constrain to better than within $100\,{\rm K}$.

\subsubsection{Deep water cycling, melting, and outgassing efficiency}\label{sec:discussion-dynamics}


Not all mantle layers participate equally in a planet's deep water cycle. We have already seen that mantles focus water storage in their transition zone NAMs. As a structural feature in the interior, the MTZ appears to be nearly ubiquitous across system compositions. Several higher-order phenomena occur around this region, which could also control how water moves across it, and how water reservoirs in the lower mantle and the shallow, ol- and opx-bearing mantle could communicate.

\medskip

\textbf{Mantle-melt density crossover}---first, where melting of mantle rock occurs, some water in NAMs will partition into the melt phase. If this melt is less dense than its associated solid residue, it will quickly rise upwards, and its water might eventually degas to the surface through volcanism---this is the scenario, for example, in the mantle beneath mid-ocean ridges on Earth. 

If instead melt is denser than the solid, water transport can only occur via solid-state convection, which is orders of magnitude slower \citep[not more than metres per year;][]{ricard_702_2015} than melt migration \citep[$\sim$30$\,{\rm m}\,{\rm yr}^{-1}$;][]{katz_physics_2022}. In general, melt is more compressible than solid mantle, meaning that the melt formed during partial melting may become negatively buoyant (i.e., denser than the solid residue) at some pressure in a planet. This precise pressure will change by a few GPa depending on the melt FeO content, but will probably be independent of Mg/Si \citep{ohtani_melting_1995, agee_compressibility_2008}. On Earth the mantle-melt density crossover pressure lies just above the MTZ, at 11--$12\,{\rm GPa}$; therefore, most of Earth's olivine-bearing mantle can efficiently participate in moving water to the surface \citep{andrault_mantle_2022}. In the exoplanet context, calculating the volume within which a mantle can produce negatively buoyant hydrous melt requires a melting model for arbitrary compositions at high pressure. Such a model is not yet available. However, if the relative amount of olivine to pyroxene were the only factor distinguishing upper mantle compositions, and if partial melting of these phases produced similar-density melts, then we might reasonably assume that most other rocky planets have an Earth-like density crossover pressure.

Knowing where (and how much) hydrous melting occurs in a mantle will matter for its outgassing. The shallow part of the UM---hosting ol and opx---may be the relevant water reservoir that supplies hotspot volcanism on Earth \citep{yang_intraplate_2020}. Our model produces a water capacity of this ol- and opx-bearing mantle of about $\sim$250--360 ppm at 1600 K, with no dependence on $M_p$ or mantle Mg/Si $\gtrsim 0.8$; i.e., the difference in water capacity between ol and pyroxenes is not substantial. This is the maximum water available to partition into any melt that forms in this region, and hence--in conjunction with the positive effect of water content on melting itself---would be a limiting factor for \ce{H2O} outgassing on oxidised stagnant lid planets \citep{guimond_low_2021}. 

\medskip

\textbf{A water saturation bottleneck}---the second phenomenon affecting deep water cycling is the presence (or absence) of a double discontinuity in NAM water saturation bracketing the MTZ. If a wadsleyite-bearing parcel is forced upwards and wad undergoes a phase transition, the water in excess of ol saturation triggers dehydration melting. The melt produced in this process would then be buoyed upwards wherever it is less dense than the residual mantle, as described above. Conversely, as material is downwelling and ring transitions to pv, the excess water exsolves and percolates back up to the MTZ: a bottleneck is created for water transport to the lower mantle \citep{bercovici_whole-mantle_2003}. This bottleneck would be bidirectional if upwelling plumes do not entrain much water and remain dry. Indeed, \citet{bercovici_whole-mantle_2003} find such a case for Earth, based on its plume velocity and geometry, and given a low H diffusivity from the ambient mantle into the plume. Plumes would also be hotter than the ambient mantle and thus have a lower water content at water saturation. On Earth, MTZ material is driven upwards as a counterflow to the downwards motion of subducting slabs; vertical motion of this layer may still occur in the absence of plate tectonics, for example, from thermal convection. 
\medskip

The framework for deep water cycling we have just described allows us to set up different scenarios where the lower mantle would or would not be an important contributor to the planetary water cycle: 
\begin{enumerate}
    \item Two-layer convection would preclude mass exchange between the lower and upper mantles \citep[e.g.,][]{tackley_mantle_1995}, so the lower mantle would not participate in the planetary water cycle.
    \item Whole-mantle convection, where upwelling produces a relatively dense melt at the MTZ, would trap water in the melt near the top of the MTZ. The layers above the MTZ would therefore be the most important for interior-surface water exchange.
    \item Whole-mantle convection, where melt produced by dehydration melting at the top of the MTZ is buoyant, and where upwelling plumes can entrain a significant amount of water as they pass through the MTZ, would entail that both the lower mantle and the upper mantle are important in the deep water cycle.
\end{enumerate}

Using the results of this study, we can make some informed guesses about the effect of bulk composition on planetary water cycling. Low-Mg/Si planets stabilise little-to-no wad, and therefore would not have a sharp water saturation discontinuity. Whilst they would still see a gradual increase in water saturation as ringwoodite starts to appear around 17--$23\,{\rm GPa}$ (Fig. \ref{fig:mgsi_modality}), even if dehydration melting occurred in this shell, it may be negatively buoyant (assuming roughly similar melt-mantle density crossovers to peridotite), so any melt would be trapped at the top of the ringwoodite-bearing layer. We speculate that such planets may have less efficient deep water cycles and drier outgassing. 



It is worth noting that not all mantles necessarily outgas. For example, sufficiently-large ocean masses can exert such a large overburden pressure that decompression melting first slows, and then shuts off entirely. Because this occurs at tens of OM for planets near $1\,M_\oplus$, however, mineralogical variations in water capacity will be relatively minor compared to the ocean masses required to stopper volcanism \citep{kite_geodynamics_2009, noack_water-rich_2016}. Nonetheless, we expect rates of water cycling to be generally negatively coupled to some degree to the mass of the surface water reservoir. 

\subsection{Possibility of SiO$_2$-rich planets}\label{sec:discussion-sio2}

Because our main study employs a high water capacity of anhydrous stv ($>1$ wt\%) based on \citeauthor{panero_hydrogen_2004}'s theoretical work (\citeyear{panero_hydrogen_2004}), even small proportions of stable silica force $w_{\rm m}$ to increase rapidly. We find this shift to occur around Mg/Si $\lesssim 0.8$ (Fig. \ref{fig:h2o_mgsi_scatter}), representing less than 4\% of Hypatia stars. Hence such compositions would be rare---if they indeed exist---yet they would suggest an entirely foreign geodynamic regime. Our phase equilibria calculations do not include bona fide hydrous stv (\ce{SiO2*H2O}), which could be stable at lower mantle pressures and increase water capacity still further \citep[see][]{nisr_large_2020}. Very water-rich mantles would be perhaps more similar, in terms of their interior structures and dynamics, to sub-Neptunes than to Earth.

However, mantle Mg/Si may not reach such extreme low ratios if fractionation processes ordinarily enrich Mg in the mantle or the bulk planet. Such processes must have occurred during Earth's formation, given that the Earth's mantle Mg/Si \citep[1.25;][]{mcdonough_composition_1995} is higher than the measured solar or chondritic values (1.04 and 1.05, respectively). In particular, Si may partition into the core during its differentiation. A fixed core Si content of 9 wt\%, for example, would shift our 2nd-percentile mantle Mg/Si from 0.72 to 0.80 given \coreeff~$ =0.113$, dropping stv and seif abundances from 5\% down to 0.4\% in the layers bearing them (although Mg/Si = 0.80 is still low enough to prevent any ol or wad from stabilising in the upper mantle). Because it is not trivial to predict the metal-silicate partitioning of Si at the core formation conditions of a given planet, previous studies have also assumed either an Si-free core as we have here \citep[e.g.,][]{dorn_can_2015,dorn_generalized_2017, dorn_interior_2018, hinkel_starplanet_2018, wang_enhanced_2019, unterborn_pressure_2019, spaargaren_influence_2020, otegi_impact_2020}, or core Si at a fixed weight percent \citep{unterborn_scaling_2016, spaargaren_compositional_2021}. Partitioning of Si from silicate melt into metallic melt does seem to increase with increasing temperature \citep{fischer_high_2015}, suggesting that larger planets with higher temperatures of core formation may have increasingly elevated mantle Mg/Si with respect to the stellar ratio, all else equal. 

The inward migration of pebbles in the protoplanetary disk may be another source of compositional discrepancy between star and planet; for example, accumulation of forsterite-rich pebbles would also deplete the planet's mantle in Si \citep{miyazaki_dynamic_2020}. 

In contrast, fractionation during magma ocean crystallisation could also decrease Mg/Si: silica enrichment may occur in the lower mantle if mantle mixing is in fact inefficient \citep{ballmer_persistence_2017}, as opposed to the homogeneous mantle oxide compositions we assume here.

\subsection{Key experimental and theoretical data needs}

Although we have tried to make the best of the available data, our predictions of absolute water capacities are necessarily approximations. However, the relative partitioning of water between phases (i.e., whether $D^j_i >1$ or $<1$) is likely to be accurate. Therefore we can put our results forward as general trends, if not rigorous quantifications, of the interior water storage capacities of individual exoplanets. To move towards the latter, we identify the most important areas where key data is missing:

\begin{enumerate}
    \item \textit{The temperature- and pressure-dependent solubility of water in pv and ppv---}together these phases compose the vast majority by volume of most rocky planet mantles. Until these water solubilities can be ascertained, upper limits on whole-mantle water storage capacities will be mostly unconstrained. These data would be the most important to have, especially for larger-mass planets where contributions from the upper mantle and MTZ are less significant; 
    \item \textit{The near-surface solidus temperatures of arbitrary mantle compositions---}these solidi set the initial potential temperature of the solid mantle, and thereby dictate the maximum initial water inventory;
    \item \textit{Mantle-melt density crossovers of arbitrary mantle compositions---}this crossover pressure determines the relevant mantle reservoir(s) in which hydrous melt would be buoyant and able to feed volcanism; and,
    \item \textit{The dependence of the water solubility of major mantle phases on FeO content---}\citet{dong_water_2022} made a start bootstrapping the few existing data points on this effect for ol, wad and ring, but not enough data is available to justify a parameterisation as of yet. 
\end{enumerate}

\section{Conclusion}

We have presented the first calculation of how much water rocky exoplanet interiors can store on the basis of mineralogy. This calculation is possible because mantle bulk compositions, input to equilibrium mineralogy models, can be approximated by host star refractory element abundances \citep{anders_solar-system_1982, thiabaud_elemental_2015, bonsor_host-star_2021}. Maximum water capacities can then be found knowing the characteristic water saturation of nominally-anhydrous minerals \citep{keppler_thermodynamics_2006}. The predictions in this work amount to a sense of change of exoplanet interior water capacity with planet mass and bulk composition. We have also considered the effects of Fe-core partitioning and mantle potential temperature.

We find that median interior water capacities in kg scale roughly as $w_{\rm m} \propto M_p^{0.69}$ for $M_p \in [0.3, 3]\,M_\oplus$. This scaling is more shallow than an uninformed scaling of planet water capacity as a constant fraction of $M_p$ because mantles concentrate their water storage in a thin shell---the mantle transition zone---at a depth that is roughly independent of planet mass. As a consequence, upper mantle water capacities reach an asymptotic limit of $\sim$1--2 OM. Larger planets are made of increasingly deeper, dry lower mantles; above $2\,M_\oplus$, the maximum water mass fractions they sequester are 300--500 ppm for almost all stellar compositions.

In the mantle transition zone, the effects of varying bulk composition (namely, the Mg/Si ratio) play out by changing the proportions of wadsleyite and ringwoodite with respect to garnet, garnet being tenfold drier than these phases. Hence greater Mg/Si generally leads to greater mantle water capacities. Lower mantle phases are less variable in both their equilibrium proportions and water saturation, which is why more massive planets show a narrower distribution of interior water mass fractions at water saturation. 


Mineral water capacities are sensitive to temperature. However, we can assume that the early potential temperature of a mantle as soon as it crystallises from the primordial magma ocean is set by its shallow solidus temperature. Further, mantles may readily inherit a water budget near water saturation \citep{tikoo_fate_2017, dorn_hidden_2021, bower_retention_2021, miyazaki_wet_2022}. Therefore, the water capacities of mantles at estimated shallow solidus temperatures ($T_p \sim 1900\,{\rm K}$) could provide a deterministic approximation to their initial water contents, irrespective of stochastic water accretion. On stagnant lid planets, this is also the hard upper limit to the ocean mass that can be outgassed during the planet's lifetime.


Ultimately, our results can inform how water is partitioned between a planet's interior and surface reservoirs. We suggest that rocky planets several times the mass of Earth may be less likely than smaller planets to maintain dry land next to oceans, motivating future work on the climate and habitability of both aqua planets and desert worlds.

\section*{Acknowledgements}

A review by Cayman Unterborn has strongly improved this manuscript. We acknowledge the support of the University of Cambridge Harding Distinguished Postgraduate Scholars Programme and the Natural Sciences and Engineering Research Council of Canada (NSERC). Cette recherche a \'et\'e financ\'ee par le Conseil de recherches en sciences naturelles et eng\'enie du Canada (CRSNG). The research shown here acknowledges use of the Hypatia Catalog Database, an online compilation of stellar abundance data as described in Hinkel et al. (2014, AJ, 148, 54), which was supported by NASA's Nexus for Exoplanet System Science (NExSS) research coordination network and the Vanderbilt Initiative in Data-Intensive Astrophysics (VIDA).

\section*{Data Availability}

The Python code used in this study is available from the corresponding author upon request.

\bibliographystyle{mnras}
\bibliography{references.bib}
\end{document}